# Classical and Quantum Shell Dynamics, and Vacuum Decay [**]

S.Ansoldi[1]
*Dipartimento di Fisica Teorica dell'Università,
Istituto Nazionale di Fisica Nucleare, Sezione di Trieste,
Strada Costiera 11, 34014-Trieste, Italy,*

A.Aurilia[2]
*Department of Physics, California State Polytechnic University,
Pomona, CA 91768, USA,*

R.Balbinot[3]
*Dipartimento di Fisica dell'Università,
Istituto Nazionale di Fisica Nucleare, Sezione di Bologna,
via Irnerio 46, 40126-Bologna, Italy,*

E.Spallucci[4]
*Dipartimento di Fisica Teorica dell'Università,
Istituto Nazionale di Fisica Nucleare, Sezione di Trieste,
Strada Costiera 11, 34014-Trieste, Italy.*

## Abstract

Following a minisuperspace approach to the dynamics of a spherically symmetric shell, a reduced Lagrangian for the radial degree of freedom is derived directly from the Einstein–Hilbert action. The key feature of this new Lagrangian is its invariance under time reparametrization. Indeed, all classical and quantum dynamics is encoded in the Hamiltonian constraint that follows from that invariance. Thus, at the classical level, we show that the Hamiltonian constraint reproduces, in a simple gauge, Israel's matching condition which governs the evolution of the shell. In the quantum case, the vanishing of the Hamiltonian (in a weak sense), is interpreted as the Wheeler–DeWitt equation for the physical states, in analogy to the corresponding case in quantum cosmology. Using this equation, quantum tunneling through the classical barrier is then investigated in the WKB approximation, and the connection to vacuum decay is elucidated.

[*]E-mail: `ansoldi@vstst0.ts.infn.it`

[†]E-mail: `aaurilia@csupomona.edu`

[‡]E-mail: `balbinot@bologna.infn.it`

[§]E-mail: `spallucci@vstst0.ts.infn.it`

## I. INTRODUCTION

According to current ideas in cosmology, in particular the inflationary scenario [1], the universe would consist of (infinitely) many self reproducing bubbles which are continuously nucleated quantum mechanically. Some of them expand and look like a Friedmann universe, others collapse to form black holes and some are connected by wormholes.

In principle, this dynamical network may exist even at a very small scale of distances: at the Planck scale, it is expected to be the manifestation of gravitational fluctuations which induce a foam-like structure on the spacetime manifold [2]. A complete analysis of the global dynamics of this structure is difficult at present, as it involves the details of a quantum theory of gravity. From our vantage point, however, this complex structure can be approximated by an ensemble of cells of spacetime, each characterized by its own constant vacuum energy density and mass. In principle, each cell may behave as a black hole, wormhole or an inflationary bubble, depending on the matching conditions on the neighboring cells. One hopes, then, that some insight in the structure of the vacuum may be gained by examining a much simpler system consisting of a self gravitating thin shell separating two spherically symmetric domains of spacetime.

Following this line of thinking, black hole, wormhole and inflationary bubble models were constructed, and their dynamics was investigated in some detail ( [3]– [9]) using the Israel matching condition between the internal and the external metric along the shell orbit in spacetime [10]:

$$\sigma_{in} R \sqrt{1 - \frac{2G_N M_{in}}{R} - \frac{\Lambda_{in}}{3} R^2 + \dot{R}^2} + $$
$$-\sigma_{out} R \sqrt{1 - \frac{2G_N M_{out}}{R} - \frac{\Lambda_{out}}{3} R^2 + \dot{R}^2} = 4\pi \rho G_N R^2 \quad . \tag{I.1}$$

In the above equation, $\Lambda_{in/out} = 8\pi G_N \epsilon_{in/out}$ are the cosmological constants associated with the internal (in) and external (out) vacuum energy, $M_{in/out}$ are the respective mass parameters, $\rho$ is the shell energy density, and $R$ is the radius of the shell. As a matter of notation, a "dot" means derivation with respect to the shell proper time, and $\sigma_{in}(\sigma_{out}) = +1$ if $R$ increases in the outward normal direction to the shell, while $\sigma_{in}(\sigma_{out}) = -1$ if $R$ decreases. Furthermore, in eq.(I.1) $\rho$ is understood as a function of $R$ to be determined from the surface stress tensor conservation law, once an equation of state relating the surface energy density and the tension of the shell is assigned.

Against this classical background, there exist at least two effective methods to approach the quantum dynamics of the shell: i) by constructing, for each model, a specific Hamiltonian operator which leads to eq.(I.1) (or, to some squared version of it) [3,5], [9,11], or, ii) by extracting from the Einstein-Hilbert action for the whole spacetime a one degree of freedom reduced action which, under variation, gives eq.(I.1) [6,8,12].

This second approach seems more rigorous and less arbitrary than the first method, inasmuch as it relies on a well established action principle in order "to run" the machinery of the Lagrangian formalism. However, one drawback common to both approaches is their dependence on the choice of the evolution parameter (internal, external, proper) with respect to which one labels the world history of the shell. The choice of time coordinate, in turn, affects the choice of a particular quantization scheme, leading, in general, to quantum theories which are not unitarily equivalent. In some cases, a particular choice of evolution parameter may even be inconsistent with the canonical quantization procedure, which is then abandoned in favor of Euclidean or path integral methods [7,8].

The root of the problem, it seems to us, is the lack of a dynamical formulation which is invariant under a general redefinition of time, and the natural way to enforce this invariance property is through an action principle, or lagrangian formalism, which encodes both the classical and quantum dynamics of the shell. On mathematical grounds, one then expects that the invariace under time reparametrization leads to a primary constraint in the corresponding Hamiltonian formulation, while on physical grounds, one must demand that the equations derived from it, be consistent with Israel's matching condition. It seems satisfying, therefore, that the above mathematical and physical requirements go hand in hand in our formulation, in the sense that the primary constraint originating from time reparametrization, automatically generates a (secondary) Hamiltonian constraint which, in a simple gauge, represents nothing but Israel's matching condition.

To summarize, then, the main purpose of this paper is to suggest a framework in which both the classical and quantum dynamics of a self gravitating spherical shell is discussed in terms of the canonical formalism of constrained systems. Thus we start from the Einstein–Hilbert action which includes the world–history of the shell and its boundary; then, by extending a procedure due to Farhi, Guth and Guven (FGG) [7], we extract from it a reduced action for the shell radial degree of freedom. The FGG–method is "extended" in the sense that, while the metric of the background manifold is fixed, the dynamical variables for the shell are chosen to be the scale factor $R(\tau)$ together with the lapse function $N(\tau)$. These variables determine the form of the intrinsic metric on the shell

$$ds_\Sigma^2 = -N(\tau)^2 d\tau^2 + R(\tau)^2 d\Omega^2 \quad , \tag{I.2}$$

where $d\Omega^2$ represents the line element of the unit 2–sphere, and $\tau$ is an arbitrary time parameter along $\Sigma$, the shell orbit in spacetime. As discussed earlier on, we insist that any redefinition of that parameter should not affect the dynamical evolution of the shell. This requirement has rather far reaching mathematical and physical implications that we discuss in the following sections. Thus, in Section II, we define our system and obtain a reduced action for the shell. In Section III, we show how the matching equation (I.1) emerges as a Hamiltonian constraint reflecting the invariance of the theory under arbitrary $\tau$-reparametrization. In Section IV, we select a special gauge in order to explore the effective dynamics of the shell with an eye on the quantum discussion which follows in Section V. There we show that the Hamiltonian constraint, which led to Israel's equation in the classical theory, now leads to the Wheeler–DeWitt equation for the quantum states of the system. This is by no means accidental, since our formulation follows closely the minisuperspace approach to quantum cosmology, so that all classical and quantum shell–dynamics is encoded in the constraint that the Hamiltonian of the system vanishes in a weak sense. Section VI is devoted to an analytic and diagrammatic study of some quantum processes which are classically forbidden. The consistency of our formulation is then tested by deriving some well established results concerning vacuum decay. Section VII ends the paper with a summary of our discussion and some concluding remarks.

In order not to obscure the logical flow of our discussion, we have assembled several important technical steps into four Appendices. Appendix A clarifies the relationship between our variational procedure and the FGG–method; in Appendix B we show how to derive the general form of the Hamiltonian; in Appendix C we derive the general expression for the nucleation coefficient in vacuum decay by calculating its defining integral in the complex plane; finally, Appendix D provides all the necessary definitions and algebraic steps to connect our results about quantum tunneling with other results already existing in the literature.

## II. THE REDUCED ACTION.

The dynamics of a generic system containing matter fields interacting with gravity is encoded in the Einstein-Hilbert action

$$S = S_g + S_m + S_B$$
$$= \frac{1}{16\pi G_N} \int_V d^4x \sqrt{g} \mathcal{R}^{(4)} + \int_V d^4x \sqrt{g} \mathcal{L}_m + \frac{1}{8\pi G_N} \int_B d^3x \sqrt{h} \mathcal{K} \quad . \tag{II.1}$$

The notation is as follows: $G_N$ stands for Newton's constant, $g$ is the determinant of the four dimensional spacetime metric, $\mathcal{R}^{(4)}$ is the corresponding Ricci scalar, $\mathcal{L}_m$ is the matter field Lagrangian density, $\mathcal{K}$ is the extrinsic curvature of the three dimensional boundary, $(B)$, of the four-dimensional region $V$ of integration, and $h$ is the determinant of the three-dimensional metric on the boundary. The presence of the surface term allows the field equations to be obtained from a variational principle in such a way that only the metric on the boundary is held fixed.

The actual physical system under consideration consists of two static, spherically symmetric, spacetimes $\mathcal{M}_1$ and $\mathcal{M}_2$ glued together along a time–like manifold $\Sigma$ which represents the world history of the shell.

Our choice for $\mathcal{L}_m$ is

$$\mathcal{L}_m = \Lambda_{in} \tag{II.2}$$

in $\mathcal{M}_1$,

$$\mathcal{L}_m = \Lambda_{out} \tag{II.3}$$

in $\mathcal{M}_2$, and

$$\mathcal{L}_m = \delta(\Sigma)\mathcal{L}_{sh} \tag{II.4}$$

along the shell, where

$$\mathcal{L}_{sh} = -\rho \quad . \tag{II.5}$$

As noted in Section I, $\Lambda_{in}$ and $\Lambda_{out}$ are the two cosmological constants, and $\rho$ is the shell energy density. Furthermore, one could easily extend the content of the action integral by adding the electromagnetic term $-(16\pi)^{-1}F_{\mu\nu}F^{\mu\nu}$ to $\mathcal{L}_m$. This addition would lead to the Reissner-Nordström-de Sitter solution in $\mathcal{M}_1$ and $\mathcal{M}_2$. However, we shall not consider this case here, as its discussion would only obscure the simplicity of the approach that we wish to illustrate.

Solving Einstein's equations in $\mathcal{M}_1$, we obtain the Schwarzschild-de Sitter solution

$$ds_1^2 = -A_{in}dT^2 + \frac{1}{A_{in}}dr^2 + r^2 d\Omega^2 \quad , \tag{II.6}$$

where

$$A_{in} = 1 - \frac{2M_{in}}{r} + \frac{\Lambda_{in}}{3}r^2 \tag{II.7}$$

and $T$, $r$, $\vartheta$, $\varphi$ are the usual Schwarzschild coordinates. Similarly, in $\mathcal{M}_2$

$$ds_2^2 = -A_{out}dt^2 + \frac{1}{A_{out}}dr^2 + r^2 d\Omega^2 \quad , \tag{II.8}$$

where

$$A_{out} = 1 - \frac{2M_{out}}{r} + \frac{\Lambda_{out}}{3}r^2 \quad . \tag{II.9}$$

The evolution of the shell, against this fixed background, is described by the radius $R(\tau)$ and by the lapse $N(\tau)$ which characterize the shell intrinsic geometry according to equation (I.2). This dynamics can be obtained by reducing the action (II.1) to a functional depending only on $N(\tau)$ and $R(\tau)$. Our procedure of reduction follows closely the discussion of ref. [7] (see also ref. [6]); one element of novelty consists in the inclusion of the lapse function $N(\tau)$, which is essential to derive the Hamiltonian constraint that one expects in a reparametrization invariant theory.

To begin with, let us specify the volume $V$ of integration. As shown in fig.1, the volume is bounded by two space-like surfaces $T = T_i$ and $T = T_f$, and by two time-like surfaces: $r = R_1$ and $r = R(\tau) > R_1$ in $\mathcal{M}_1$, and by $t = t_i$, $t = t_f$, $r = R(\tau)$ and $r = R_2 > R(\tau)$ in $\mathcal{M}_2$. $T_1$, $t_1$, $T_2$, $t_2$, $R_1$, $R_2$ are constants. Furthermore, we assume, for the moment, that $V$ lies in a region of spacetime where the Killing vectors of the metric (II.6-II.8), $\partial_T$ and $\partial_t$ respectively, are both time-like. Later on, we shall give a supplementary rule for the case in which the above condition is not satisfied.

FIGURES

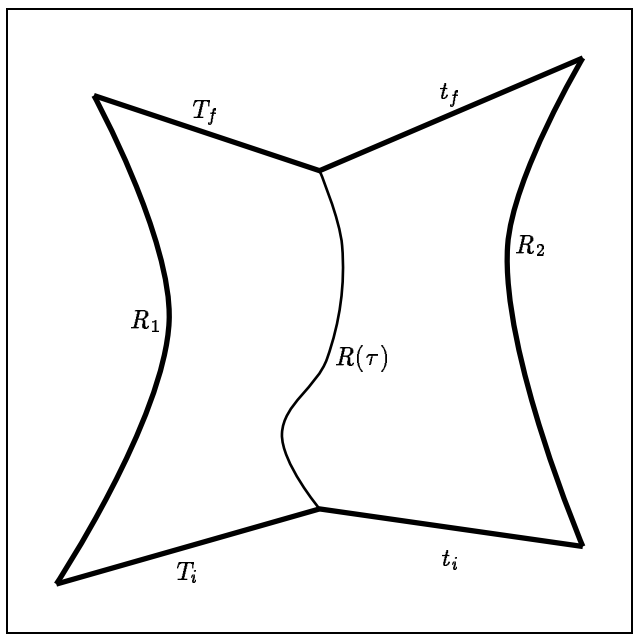

FIG. 1. Graphical representation of the integration volume with two dimensions suppressed: $T_i$ ($T_f$) and $t_i$ ($t_f$) label the initial (final) surfaces which are the space–like boundaries of the integration volume. $R_1$ and $R_2$ represent the time–like boundaries, and $R(\tau)$ parametrizes the shell radius which separates the interior domain from the exterior one.

Evaluating the induced metric on both side of the shell, and comparing with eq.(I.2), we find the implicit dependence of the Schwarzschild time variables on the parameter $\tau$,

$$\frac{dt}{d\tau} = \dot{t} = \frac{\beta_{out}}{A_{out}} \quad , \tag{II.10}$$

where

$$\beta_{out} = \sqrt{N^2 A_{out} + \dot{R}^2} \tag{II.11}$$

and

$$\frac{dT}{d\tau} = \dot{T} = \frac{\beta_{in}}{A_{in}} \quad , \tag{II.12}$$

where

$$\beta_{in} = \sqrt{N^2 A_{in} + \dot{R}^2} \quad . \tag{II.13}$$

Note that the $\beta$–functions defined above contain a sign ambiguity [7], which is resolved by requiring that $\beta$ be positive (negative) if the outer normal to $\Sigma$ points toward increasing (decreasing) values of $r$.

Our immediate objective, for the remainder of this section, is to obtain the reduced form of the action integral, and we shall do so by discussing each individual term separately. The explicit form of $S_m^\Sigma$, the contribution of the shell to $S_m$, is given by

$$S_m^\Sigma = 4\pi \int_{\tau_i}^{\tau_f} \rho\left(R\left(\tau\right)\right) R(\tau)^2 d\tau \tag{II.14}$$

supplemented by an equation of state relating the energy density $\rho$ to the tangential pressure $p$. Such a relation is usually obtained from the conservation equation for the shell stress–energy tensor,

$$\dot\rho = -2(\rho - p)\frac{\dot R}{R} \quad, \tag{II.15}$$

and we shall discuss some simple special cases of this equation later on in the text.

Next, in order to calculate $S_g^\Sigma$, the contribution of the shell to $S_g$ in the action integral, we first introduce Gaussian normal coordinates near the shell

$$ds^2 = g_{\tau\tau}\left(\tau, \eta\right) d\tau^2 + d\eta^2 + r\left(\tau, \eta\right)^2 d\Omega^2 \quad. \tag{II.16}$$

As usual, the triplet $(\tau, \vartheta, \varphi)$ specifies a point on the shell, while $\eta$ represents the geodesic distance off the shell. One has $\eta > 0$ in $\mathcal{M}_2$, and $\eta < 0$ in $\mathcal{M}_1$. Furthermore,

$$g_{\tau\tau}\left(\tau, 0\right) = -N(\tau)^2 \tag{II.17}$$
$$r\left(\tau, 0\right) = R(\tau) \quad. \tag{II.18}$$

One can now start the evaluation of $S_g^\Sigma$:

$$S_g^\Sigma = \frac{1}{16\pi G_N} \int_{-\epsilon}^{+\epsilon} d\eta \int_{\tau_i}^{\tau_f} d\tau \int d\vartheta d\varphi \sqrt{g}\, \mathcal{R}^{(4)} \quad. \tag{II.19}$$

Here $\epsilon$ is an arbitrary small positive number, which at the end of the calculation is set to zero. The Ricci scalar $\mathcal{R}^{(4)}$ can be expressed as

$$\mathcal{R}^{(4)} = \mathcal{R}^{(3)} - \left(K_{ij}K^{ij} + K^2\right) - 2\frac{\partial K}{\partial \eta} \tag{II.20}$$

where $K_{ij}$ stands for the extrinsic curvature of the hyper-surface of constant $\eta$, and $\mathcal{R}^{(3)}$ is the Ricci scalar constructed from the three-dimensional metric $g_{ij}$ on the hyper-surface of constant $\eta$. The relation between $K_{ij}$ and the normal derivative of the three-metric is simply

$$K_{ij} = \frac{1}{2}\frac{\partial g_{ij}}{\partial \eta} \quad. \tag{II.21}$$

Evaluating these quantities, one eventually obtains

$$K_{\vartheta\vartheta} = \frac{K_{\varphi\varphi}}{\sin^2 \vartheta} = \frac{R\beta_{in}}{N} \quad \text{for} \quad \eta = -\epsilon \tag{II.22}$$

$$K_{\vartheta\vartheta} = \frac{K_{\varphi\varphi}}{\sin^2 \vartheta} = \frac{R\beta_{out}}{N} \quad \text{for} \quad \eta = +\epsilon \tag{II.23}$$

and

$$K_{\tau\tau} = -2\frac{\ddot{R}N}{\beta_{in}} - \frac{N^2}{\beta_{in}}\frac{dA_{in}}{dR} + 2\frac{\dot{R}\dot{N}}{\beta_{in}} \quad \text{for} \quad \eta = -\epsilon \tag{II.24}$$

$$K_{\tau\tau} = -2\frac{\ddot{R}N}{\beta_{out}} - \frac{N^2}{\beta_{out}}\frac{dA_{out}}{dR} + 2\frac{\dot{R}\dot{N}}{\beta_{out}} \quad \text{for} \quad \eta = +\epsilon. \tag{II.25}$$

All of the above leads to the following expression of $S_g^\Sigma$

$$S_g^\Sigma = \frac{1}{2G_N}\int_{\tau_i}^{\tau_f} d\tau \left[2\beta R + \frac{R^2}{\beta}\left(\ddot{R} + \frac{N^2}{2}\frac{dA}{dR} - \frac{\dot{R}\dot{N}}{N}\right)\right]_{out}^{in} . \tag{II.26}$$

In eq.(II.26) the notation $[\ldots]_{out}^{in}$ means, as usual, $[\ldots]]_{in} - [\ldots]]_{out}$.

The contributions of $\mathcal{M}_1$ and $\mathcal{M}_2$ to $S_g$, which we denote by $S_g^1$ and $S_g^2$ respectively, are easily evaluated since

$$\mathcal{R}_{in/out}^{(4)} = 4\Lambda_{in/out} . \tag{II.27}$$

This reduces the form of $S_g^{1\,(2)}$ to a volume integral of the same type as $S_m^{1\,(2)}$.

Finally, the surface term in the action integral, which eliminates the second derivative term $\ddot{R}$ present in eq.(II.26), can be written as

$$S_B = \frac{1}{8\pi G}\int d^3x\sqrt{h}\nabla_\mu n^\mu , \tag{II.28}$$

where $n^\mu$ is the unit normal to the boundary three–surface, and $h$ is the determinant of the metric of the three–surface. Again, following the same steps as in ref. [7], one obtains

$$\begin{aligned}S_B^\Sigma &= -\frac{1}{2G_N}\int_{\tau_i}^{\tau_f} d\tau \frac{d}{d\tau}\left\{R^2\left[\tanh^{-1}\left(\frac{\dot{R}}{\beta}\right)\right]_{out}^{in}\right\} \\ &= -\frac{1}{2G_N}\left[R^2\dot{R}\tanh^{-1}\left(\frac{\dot{R}}{\beta}\right) + \right. \\ &\quad \left. +\frac{R^2}{A}\left(\frac{A\ddot{R}}{\beta} - \frac{\dot{R}^2}{2\beta}\frac{dA}{dR} - \frac{A\dot{R}\dot{N}}{N\beta}\right)\right]_{out}^{in} .\end{aligned} \tag{II.29}$$

The end result of the above calculations is the following expression of the action integral

$$S = \int_{\tau_i}^{\tau_f} L d\tau + \text{ boundary terms } , \tag{II.30}$$

where

$$L = \left[-\frac{R\dot{R}}{G_N}\tanh^{-1}\left(\frac{\dot{R}}{\beta}\right)\right]_{out}^{in} + \frac{R}{G_N}\left(\beta_{in} - \beta_{out} - \kappa RN\right) \tag{II.31}$$

and $\kappa \equiv 4\pi\rho G_N$. The additional "boundary terms" in eq.(II.30) collectively refer to contributions which are simply proportional to $\int_{t_1}^{t_2} dt$ or $\int_{T_1}^{T_2} dt$. Their explicit form is irrelevant for our purposes. Indeed, since the intrinsic dynamics of the shell depends only on the interior and exterior geometry, both of which are fixed, and not on the spacetime volume $V$ chosen, we define the shell effective action by subtracting the boundary terms in eq.(II.30) [6]. Therefore, the promised form of the reduced, or effective, action is given by

$$S_{eff} = \int_{\tau_i}^{\tau_f} L d\tau \quad . \tag{II.32}$$

As a consistency check, in the next section we shall verify that varying this reduced action yields the Israel equation of motion (I.1) for the shell.

### III. EQUATION OF MOTION

The purpose of this section is twofold: i) to set up the Hamiltonian formalism in preparation of the quantization procedure, and ii) to show that the equations of motion of the shell are equivalent to the matching condition (I.1). This section is supplemented by Appendix A in which we connect our canonical formalism to the FGG–approach.

Presently, our first step is to define the conjugate momenta corresponding to the dynamical variables $N$ and $R$:

$$P_N = \frac{\partial L}{\partial \dot{N}} = 0 \tag{III.1}$$

$$P_R = \frac{\partial L}{\partial \dot{R}} = -\frac{R}{G_N} \left[ \tanh^{-1}\left(\frac{\dot{R}}{\beta}\right) \right]_{out}^{in} \quad . \tag{III.2}$$

From the above equations, we then obtain the form of the Hamiltonian for our system

$$H = P_R \dot{R} - L = -\frac{R}{G_N} \left( \beta_{in} - \beta_{out} - \kappa N R \right) \quad . \tag{III.3}$$

Since $L$ is independent of $\dot{N}$, it follows that the corresponding conjugate momentum vanishes identically (eq.(III.1)). That relation represents a primary constraint which reflects the invariance of $S_{eff}$ under any $\tau$–reparametrization which maps the boundary $(\tau_i, \tau_f)$ into itself: for $\tau \to \tilde{\tau} = \tilde{\tau}(\tau)$ with $\tilde{\tau}_i = \tau_i$ $\tilde{\tau}_f = \tau_f$, we have $S_{eff} \to S_{eff}$, as it is easily seen if one keeps in mind the tensorial character of $N^2 = -g_{00}$. This primary constraint, in turn, generates a secondary constraint, namely the vanishing of the Hamiltonian (III.3). Indeed, the equations of motion of the shell are obtained by requiring that $S_{eff}$ be stationary under variation of the functions $N(\tau)$ and $R(\tau)$ subject to the overall condition that they vanish at the (fixed) boundaries $\tau_i$ and $\tau_f$: the variation of $S_{eff}$ gives

$$\delta S_{eff} = \int_{\tau_i}^{\tau_f} \left\{ \left[ \frac{\partial L}{\partial R} - \frac{d}{d\tau}\left(\frac{\partial L}{\partial \dot{R}}\right) \right] \delta R(\tau) + \frac{\partial L}{\partial N} \delta N(\tau) \right\} d\tau \quad . \tag{III.4}$$

Then, demanding that $\delta S_{eff}$ vanishes, yields two independent relations. The first

$$\frac{\partial L}{\partial N} = \frac{R}{G_N N} [\beta_{in} - \beta_{out} - \kappa N R] = -\frac{H}{N} = 0 \tag{III.5}$$

is not a true equation of motion, since $\ddot{R}$ does not appear in it. Rather, Eq.(III.5) represents a constraint on the physically allowed states of our system, and since $N(\tau) \neq 0$, it implies the (weakly) vanishing of the Hamiltonian

$$H = 0 \quad . \tag{III.6}$$

On the other hand, the second equation following from the requirement that $\delta S_{eff}$ vanishes,

$$\frac{\partial L}{\partial R} - \frac{d}{d\tau}\left(\frac{\partial L}{\partial \dot{R}}\right) = 0 \tag{III.7}$$

can be rearranged in the form

$$\frac{d}{d\tau}\left(\frac{H}{N}\right) = 0 \quad , \tag{III.8}$$

which tells us, on account of eq.(III.6), that the constraint $H = 0$ is preserved in ($\tau$) time, and that the lapse function is completely arbitrary, as expected. As a matter of fact, all classical and quantum evolution of the shell is encoded in the Hamiltonian constraint (III.6). However, the dynamics of the shell cannot be fully specified without fixing a gauge in our reparametrization invariant formulation, and this invariance is reflected in the arbitrariness of $N(\tau)$. A natural choice is the gauge $N \equiv 1$, which corresponds to selecting $\tau$ as proper time along the world history of the shell. In this gauge, the constraint reads

$$R\left(\beta_{in} - \beta_{out}\right) = 4\kappa R^2 \quad , \tag{III.9}$$

which is just the matching equation [10], now playing the role of a Hamiltonian constraint describing the classical motion of the shell.

## IV. EFFECTIVE DYNAMICS

As discussed in the Introduction, shell–dynamics covers a wide range of physical situations, from gravitational collapse into black holes, to wormholes and inflationary bubbles. Several analytical and graphical methods have been proposed to deal with the effective dynamics of a shell [13]. From our present perspective, the essential idea and technical steps can be summarized thus: let us rewrite the Hamiltonian constraint, taking explicitly into account the sign multiplicity of the $\beta$ functions,

$$R\left(\sigma_{in}\beta_{in} - \sigma_{out}\beta_{out}\right) = \kappa R^2 \quad . \tag{IV.1}$$

Squaring this expression, we obtain

$$\sigma_{in} = \text{Sgn}\left[\kappa\left(A_{in} - A_{out} + \kappa^2 R^2\right)\right] \tag{IV.2}$$

$$\sigma_{out} = \text{Sgn}\left[\kappa\left(A_{in} - A_{out} - \kappa^2 R^2\right)\right] \quad . \tag{IV.3}$$

These relations determine the signs of the $\beta$ functions along the shell trajectory.

Squaring twice the constraint (IV.1), one arrives at a simple expression for the equation of motion of the shell

$$\dot{R}^2 + V(R) = 0 \quad , \tag{IV.4}$$

where

$$\begin{aligned} V(R) &= -\frac{(A_{in} - A_{out} - \kappa^2 R^2)^2}{4\kappa^2 R^2} + A_{out} \\ &= -\frac{(A_{in} - A_{out} + \kappa^2 R^2)^2}{4\kappa^2 R^2} + A_{in} \\ &= -\frac{\left[(A_{in} + A_{out} - \kappa^2 R^2)^2 - 4 A_{in} A_{out}\right]}{4\kappa^2 R^2} \quad . \end{aligned} \tag{IV.5}$$

Remarkably, the radial evolution of the shell is governed by an equation which is equivalent to the classical energy equation of a unit mass particle moving in a potential $V(R)$ with zero energy: classically allowed paths require $\dot{R}^2 > 0$, i.e. $V(R) < 0$. Turning points correspond to $\dot{R} = 0$, i.e., $V(R) = 0$. When $V(R) > 0$ one has $\dot{R}^2 < 0$, and one speaks of a forbidden (or Euclidean) trajectory. Thus, using eq.(IV.4), one can analyze the motion of the shell by specifying, case by case, the interior and exterior geometry ($A_{in}$, $A_{out}$), as well as the shell matter content ($\rho(R)$) [13]. The analysis is facilitated by plotting, on an energy diagram, the potential curve $V(R)$, together with the horizon curves ($A = 0$) and the curves corresponding to the vanishing of the $\beta$ functions. Presently, we simply make some general observations following from the analysis of eqs.(IV.2-IV.5):

1. turning points (i.e. zeros of the potential) exist only if $A_{in} > 0$ and $A_{out} > 0$;

2. if $A > 0$, $\beta$ can change sign only along a forbidden path; if $A < 0$, $\beta$ can change sign along a classical path.

Furthermore, in the course of the above analysis, one must keep in mind that the equations of motion were obtained by considering trajectories which lie in regions where both Killing vectors $\partial_T$ and $\partial_t$ are time-like (i.e., $A_{in} > 0$ and $A_{out} > 0$). When the above conditions are not satisfied, our choice of integration volume is no longer appropriate, since the time variable ($t$ or $T$) becomes a space-like coordinate, and does not have a regular behavior at the horizons (i.e., when $A = 0$). With hindsight, however, since our effective action correctly leads to Israel's equation of motion $H = 0$, and makes no reference to the integration volume, we assume that the expression for $S_{eff}$ is valid for every shell trajectory when supplemented by a working prescription suggested in Ref. [7]: when one of the factor $A$ becomes negative, i.e. when $\dot{R}/\beta > 1$, the replacement $\tanh^{-1}(\dot{R}/\beta) \to \tanh^{-1}(\beta/\dot{R})$ should be understood in eq.(II.31).

The dynamics of the shell can also be analyzed directly in terms of the Hamiltonian (III.3)

$$H = -\frac{R}{G_N}(\beta_{in} - \beta_{out} - \kappa N R) \quad . \tag{IV.6}$$

However, the usefulness of the above expression is limited by the fact that it is given in implicit form, i.e., it contains the velocity $\dot{R}$ which, in turn, is a function of the phase space variables $(R, P_R)$ obtained by inverting eq.(III.2)

$$P_R = -\frac{R}{G_N}\left[\tanh^{-1}\left(\frac{\dot{R}}{\beta}\right)\right]_{out}^{in} . \tag{IV.7}$$

Therefore, our immediate task is to recast the Hamiltonian in a form which is more amenable to physical applications, an example of which will be discussed at the end of this section.

From equation (IV.7), it follows

$$\frac{\partial P_R}{\partial \dot{R}} = -\frac{R}{G_N}\left(\frac{1}{\beta_{in}} - \frac{1}{\beta_{out}}\right) = \frac{\kappa N R^2}{G_N \beta_{in} \beta_{out}} , \tag{IV.8}$$

where the Hamiltonian constraint has been used. If $\beta$ changes sign, then $P_R$ is not a monotonic function of $\dot{R}$, and its inverse is defined only in those intervals in which eq.(IV.8) is either positive or negative. As shown before, $\beta$ can change sign along a classical path when $A < 0$ (in this case $\tanh^{-1}(\beta/\dot{R})$ vanishes), or on a forbidden path, when $A > 0$. In a forbidden region, $\dot{R}$ is imaginary, in which case we define the Euclidean momentum by analytic continuation of eq.(IV.7)

$$P_R^E = -\frac{R}{G_N}\left[\tan^{-1}\left(\frac{\dot{R}_E}{\beta_E}\right)\right]_{out}^{in} , \tag{IV.9}$$

with

$$\frac{\pi}{2} < \arctan\left(\frac{\dot{R}_E}{\beta_E}\right) < \frac{\pi}{2} , \tag{IV.10}$$

and $\beta_E \equiv \sqrt{N^2 A - \dot{R}_E^2}$. Evidently, the above definition is not unique. For instance, the choice of the interval $[-\pi, 0]$ (as in ref. [7]), leads to a non-vanishing $P_E$ at turning points when the $\beta$'s have opposite sign there. On the other hand, our choice (IV.10) has the disadvantage that $P_E$ is not continuous when $\beta$ changes sign, i.e. $\tanh^{-1}(\beta/\dot{R}) \to \pi/2$ for $\beta_E \to 0^+$, whereas $\tanh^{-1}(\beta/\dot{R}) \to -\pi/2$ for $\beta_E \to 0^-$. In terms of spacetime diagrams, the vanishing of $\beta_E$ corresponds to an Euclidean trajectory which jumps from a region of the Penrose diagram where the normal to the shell points, say, towards increasing values of $r$, to a region where the normal points toward decreasing values of $r$. Therefore, we will limit our considerations only to those systems for which the $\beta$ functions have a definite sign. Presumably, some singular cases such as the quantum mechanical nucleation of wormholes, which involve Euclidean trajectories along which $\beta$ changes sign, may be dealt with in terms of pseudo-manifolds [7], or degenerate vierbeins [14].

Coming back to eq.(IV.7), we can rewrite it as

$$\cosh\left(\frac{G_N P_R}{R}\right) = \frac{\beta_{in}\beta_{out} - \dot{R}^2}{N^2\sqrt{A_{in}A_{out}}} . \tag{IV.11}$$

Then, combining this equation with the implicit form of the Hamiltonian (IV.6), one can eliminate the $\beta$'s and $\dot{R}$, obtaining finally the promised form of the Hamiltonian

$$H = \kappa N R^2 - \frac{NR}{G_N} \left[ A_{in} + A_{out} - 2\sigma_{in}\sigma_{out}\sqrt{A_{in}A_{out}} \cosh\left(\frac{G_N P}{R}\right) \right]^{\frac{1}{2}} \quad . \tag{IV.12}$$

This expression actually corresponds to the case $A_{in} > 0$, $A_{out} > 0$. One can proceed in a similar fashion in the other cases. A more complete account of the results is given in Appendix B. Here, however, just to give a sense of the applicability of this Hamiltonian formulation, we mention a particular form of the Hamiltonian (IV.12) which has already appeared in the literature [4]. It corresponds to a shell of dust ($p = 0$, i.e., $\rho \sim R^{-2}$) separating a Minkowski (interior) spacetime ($A_{in} = 1$, $\sigma_{in} = +1$), from a Schwarzschild (exterior) one ($A_{out} = 1 - 2G_N M_{out}/R$, $\sigma_{out} = +1$). In this case, the matching equation becomes (in the gauge $N = 1$)

$$\frac{R}{G_N}(\beta_{in} - \beta_{out}) = m \quad , \tag{IV.13}$$

where $m$ is a constant representing the rest mass of the shell. In the light of our present formulation, one recognizes the l.h.s. of eq.(IV.13) as the Hamiltonian of the shell,

$$H = \frac{R}{G_N} \left[ 2 - \frac{2G_N M_{out}}{R} - 2\sqrt{1 - \frac{2G_N M_{out}}{R}} \cosh\left(\frac{G_N P_R}{R}\right) \right]^{\frac{1}{2}} \quad . \tag{IV.14}$$

This expression coincides, up to a constant term and a sign, with the special form that our eq.(IV.12) takes under the above hypotheses ($A_{in} = 1$, $A_{out} = 1 - 2G_N M_{out}/R$, $\sigma_{in} = \sigma_{out} = +1$, $\rho \sim R^{-2}$, $N = 1$), and leads to the same dynamics. In Ref.( [5]), however, a rather different, and equally arbitrary construction is performed. In this new interpretation of the model, (see also ref. [9,11]), after squaring eq.(IV.13), one identifies $M_{out}$, instead of $m$, as the numerical value of the shell proper Hamiltonian which now reads

$$H = m\cosh\left(\frac{P_R}{m}\right) - \frac{G_N M_{out}^2}{2R} \quad . \tag{IV.15}$$

Note the inequivalence of the two effective Hamiltonians (IV.14) and (IV.15) which underscores the arbitrariness of the above constructions in the absence of a coherent and unified approach to shell dynamics.

## V. QUANTUM MECHANICS

Since the classical dynamics was formulated in terms of the well established Lagrangian formalism, it seems natural to quantize the system by interpreting the phase space variables $R$ and $P_R$ as the basic observables, or quantum operators acting on a Hilbert space $\mathcal{H}$. Then the correspondence principle

$$P_R \to \hat{P}_R = -i\hbar \frac{1}{R}\frac{\partial}{\partial R} R \tag{V.1}$$

$$R \to \hat{R} = R \tag{V.2}$$

leads to the canonical commutation relation: $\left[\hat{R}, \hat{P}_R\right] = i\hbar$. The Hamiltonian constraint of the classical theory is then imposed on the quantum states

$$\hat{H}(P_R, R)|\Psi\rangle = 0 \quad , \tag{V.3}$$

thus selecting a linear physical subspace of $\mathcal{H}$.

Equation (V.3) is nothing but the Wheeler–DeWitt equation for the quantum theory of "shell–dynamics".

Unfortunately, this straightforward procedure leads to a non local form of the Hamiltonian (IV.12) which, as a quantum operator, is plagued by ordering ambiguities. The point of fact, then, is that any manipulation of that Hamiltonian has a degree of arbitrariness that goes with it. For instance, a suggestion was made in ref. [4,5], whereby, after a canonical transformation, eq.(V.3) is transformed into a finite difference equation (the representation used there differs, however, from eqs.(V.1-V.2).

Alternatively, extrapolating the result of Section IV, namely, that the effective dynamics of a shell is simulated by the motion of a classical particle in a given potential, one may interpret the relation (V.3), as the stationary equation for a quantum particle of rest mass $m$. As a consistency check on this interpretation, let us consider the limit, as $G_N \to 0$, of the Hamiltonian $\hat{H}$ corresponding to the case, mentioned earlier, of a shell of dust separating an interior Minkowski spacetime from an exterior Schwarzschild one. In this case, we have $A_{in} = 1$, $A_{out} = 1 - 2G_N M/R$, $\kappa R^2 = m = const$, $\sigma_{in} = \sigma_{out} = +1$, and the corresponding Hamiltonian operator is given by

$$\hat{H} = m - \frac{2R}{G_N}\left[1 - \frac{G_N M}{R} - \sqrt{1 - \frac{2G_N M}{R}}\cosh\left(\frac{G_N \hat{P}_R}{R}\right)\right]^{\frac{1}{2}} \tag{V.4}$$

with $\hat{P}_R$ given by equation(V.1). Expanding in power of $G_N$, with a suitable choice of ordering, one obtains

$$\hat{H} \sim m - \frac{R}{G_N}\left[\frac{G_N^2 M^2}{R^2} - \frac{G_N^2 \hat{P}_R^2}{R^2} + \mathcal{O}\left(G_N^3\right)\right]^{\frac{1}{2}} = m - \left[M^2 - \hat{P}_R^2\right]^{\frac{1}{2}} \quad . \tag{V.5}$$

Therefore, in the limit $G_N \to 0$, the Wheeler–DeWitt equation reduces to

$$m\Psi = \sqrt{M^2 - \hat{P}_R^2}\,\Psi \quad . \tag{V.6}$$

Squaring the above relation, yields the Klein-Gordon equation for an S-wave particle of rest mass $m$ and energy (ADM energy) M, which seems to be a natural result for a spherical shell of dust, once the gravitational interaction has been switched off.

Coming back to the Wheeler-De Witt equation (V.3), although its full solution is presently out of reach, a special class of solutions exists in the literature representing a sufficiently significant sample of the shell quantum dynamics. These are the WKB solutions of the form

$$\Psi = e^{\frac{i}{\hbar}S_{eff}} \quad . \tag{V.7}$$

However, rather then reviewing the explicit construction of these solutions (see ref. [15]), in the next section we will discuss the application of the WKB formalism to some quantum tunneling processes which are forbidden according to the classical dynamics discussed in Section IV.

## VI. QUANTUM TUNNELING AND VACUUM DECAY

Let us consider, for example, the classical motion of a shell in the potential of Fig.2.

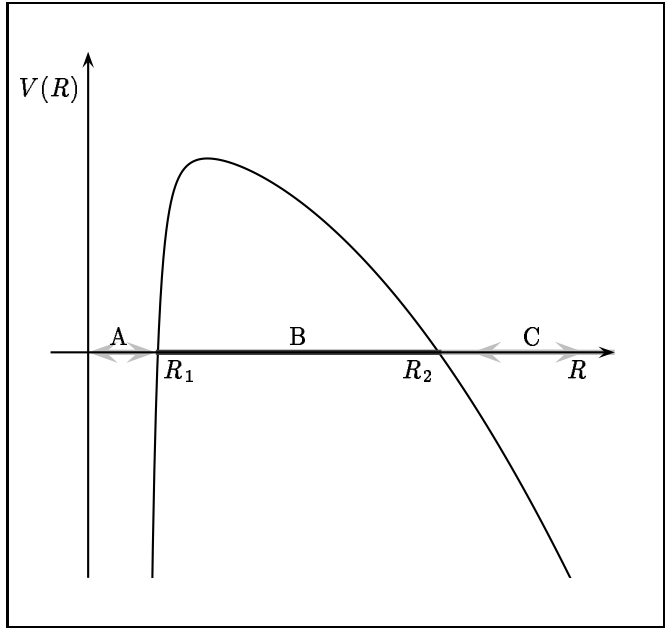

FIG. 2. Graphical representation, in a generic case, of the effective potential described by equation (IV.5). Typically, there are three distinct trajectories. Two of them are classically allowed: (A) bounded, and (C) unbounded. Trajectory (B) is classically forbidden and corresponds to quantum mechanical tunneling through the potential barrier. The exponential of the action evaluated along B gives the WKB approximation to the tunneling amplitude from A to C.

The classically allowed trajectories for the shell are: $A$ (bounded), i.e., the shell starts at $R = 0$, expands reaching a maximum radius at $R_1$ and then re-collapses to $R = 0$, or $C$ (unbounded), i.e., the shell starts with unbounded radius, contracts reaching a minimum radius at $R_2$ and then re-expands to infinity. The quantum mechanical amplitude for the shell to tunnel from one classically allowed trajectory $(A)$ to the other $(C)$, is proportional to the exponential of the integral of the Euclidean momentum $P_E$ calculated along the Euclidean trajectory $B$ which interpolates between the two classical ones, namely

$$P \sim e^{-\frac{B}{\hbar}} \quad , \tag{VI.1}$$

where

$$B = 2 \int_{R_1}^{R_2} |P_E(R)| \, dR \quad . \tag{VI.2}$$

The Euclidean momentum $P_E$ is given by eq.(IV.9). Along the Euclidean trajectory, using the equation of motion $H = 0$, that equation can be rewritten as

$$P_E(R) = -\frac{R}{G_N} \arccos\left(\frac{A_{in} + A_{out} - \kappa^2 R^2}{2\sigma_{in}\sigma_{out}\sqrt{A_{in}A_{out}}}\right) \tag{VI.3}$$

with

$$-\pi < \arccos < 0 \quad . \tag{VI.4}$$

Inserting $P_E(R)$ in eq.(VI.2), integrating by parts, taking into account the vanishing of $P_E$ at the turning points $R_1$ and $R_2$, one finally obtains

$$B = \frac{1}{G_N} \int dR R^2 \left[ \frac{4\kappa^2 R^2 A_{in} A_{out} + (A'_{in} + A'_{out}) A_{in} A_{out}}{2 A_{in} A_{out} \sqrt{4 A_{in} A_{out} - (A_{in} + A_{out} - \kappa^2 R^2)^2}} \right.$$
$$\left. - \frac{A_{in}^2 A'_{out} + A'_{in} A_{out}^2 - \kappa^2 R^2 (A'_{in} A_{out} + A_{in} A'_{out})}{2 A_{in} A_{out} \sqrt{4 A_{in} A_{out} - (A_{in} + A_{out} - \kappa^2 R^2)^2}} \right] \quad . \tag{VI.5}$$

where the symbol " ' " means differentiation with respect to $R$. In order to check the consistency of our formulation, in the remainder of this section we shall revisit the problem of vacuum decay and the influence of gravity on it. This subject was discussed in a seminal paper by Coleman and De Luccia [16], and also by Parke [17], using simple semi-classical and geometrical arguments. Our intent, then, is to show how their results follow from the dynamical framework discussed so far.

The system under consideration consists of two de Sitter domains joined along the spherical shell $\Sigma$, which is characterized by a constant surface tension $\rho$. In other words, we assign $\rho = p$ as equation of state; then, equation (II.15) implies $\rho = const.$, which is assumed to be positive. Therefore, the form of the metric in the interior and exterior regions is

$$ds_1^2 = -\left(1 - \frac{\Lambda_{in}}{3} r^2\right) dT^2 + \left(1 - \frac{\Lambda_{in}}{3} r^2\right)^{-1} dr^2 + r^2 d\Omega^2 \tag{VI.6}$$

and

$$ds_2^2 = -\left(1 - \frac{\Lambda_{out}}{3} r^2\right) dt^2 + \left(1 - \frac{\Lambda_{out}}{3} r^2\right)^{-1} dr^2 + r^2 d\Omega^2 \quad . \tag{VI.7}$$

The Hamiltonian of the shell is obtained from eq.(III.3)

$$H = -\frac{R}{G_N} \left[ \sigma_{in} \sqrt{1 - \frac{\Lambda_{in}}{3} R^2 + \dot{R}^2} - \sigma_{out} \sqrt{1 - \frac{\Lambda_{out}}{3} R^2 + \dot{R}^2} - \kappa R \right] \quad . \tag{VI.8}$$

In order to simplify the notation, let us introduce two parameters $\alpha$ and $\gamma$ such that

$$\kappa^2 \alpha = \frac{4}{3} \Lambda_{in} \tag{VI.9}$$

$$\kappa^2 \gamma = \frac{\Lambda_{out} - \Lambda_{in}}{3} + \kappa^2 \tag{VI.10}$$

and rescale the shell radius and proper time by a factor $\kappa$

$$\xi = \kappa R \tag{VI.11}$$

$$\bar{\tau} = \kappa \tau \quad . \tag{VI.12}$$

Then the coefficients of the metric become

$$A_{in} = 1 - \frac{\alpha}{4}\xi^2 \qquad (VI.13)$$

$$A_{out} = 1 - \left(\gamma - 1 + \frac{\alpha}{4}\right)\xi^2 \qquad (VI.14)$$

and the equation of motion, after dividing by $R$, can be rewritten as

$$\sigma_{in}\sqrt{1 - \frac{\alpha}{4}\xi^2 + \left(\frac{d\xi}{d\bar{\tau}}\right)^2} - \sigma_{out}\sqrt{1 - \left(\gamma - 1 + \frac{\alpha}{4}\right)\xi^2 + \left(\frac{d\xi}{d\bar{\tau}}\right)^2} = \xi \quad . \qquad (VI.15)$$

After squaring eq.(VI.15) we arrive at

$$\left(\frac{d\xi}{d\bar{\tau}}\right)^2 + V(\xi) = 0 \quad , \qquad (VI.16)$$

where

$$V(\xi) = 1 - \left(\frac{\xi}{\xi_N}\right)^2 \qquad (VI.17)$$

and

$$\xi_N = \frac{2}{\sqrt{\alpha + \gamma^2}} \quad . \qquad (VI.18)$$

Equation(VI.16) is integrated with the initial condition $\xi(\bar{\tau} = 0) = \xi_N$, so that

$$\xi = \xi_N \cosh\left(\frac{\bar{\tau}}{\xi_N}\right) \quad . \qquad (VI.19)$$

The potential $V(\xi)$ is represented in fig.3.

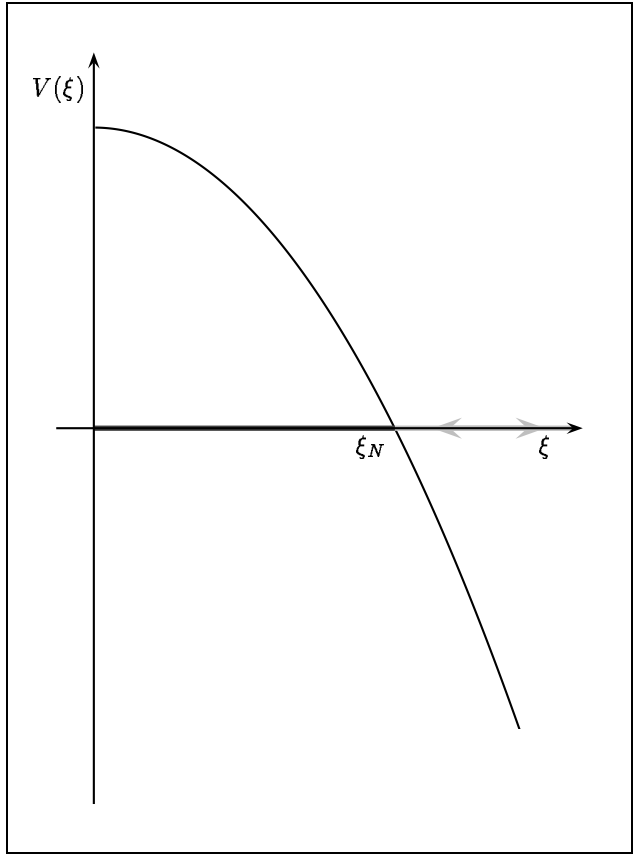

FIG. 3. Graphical representation of the effective potential $V(\xi)$ for the situation in which the "in" and "out" geometries are of de Sitter type. $\xi_N$ will be referred to as the "nucleation radius". The diagram displays only the unbounded trajectory, since the bounded one is degenerate ($R \equiv 0$). Note that only for $\sigma_{in} = \sigma_{out}$ this is a classical solution of eq.(IV.5).

The classical motion corresponds to a shell (separating the two de Sitter spaces) which starts at $\bar{\tau} = -\infty$ with an infinite radius, contracts to a minimum radius $\xi_N$ at $\bar{\tau} = 0$, and then expands again to infinity. From the equation of motion, taking into account eqs.(IV.2-IV.3), we determine the sign of the $\beta$'s as follows

$$\sigma_{in} = \text{Sgn}\left(\frac{\Lambda_{out} - \Lambda_{in}}{3} + \kappa^2\right) \tag{VI.20}$$

$$\sigma_{out} = \text{Sgn}\left(\frac{\Lambda_{out} - \Lambda_{in}}{3} - \kappa^2\right) \quad . \tag{VI.21}$$

For the purpose of illustrating our method, consider the case $\Lambda_{out} > \Lambda_{in}$, thus fixing $\sigma_{in} = +1$. Then, from eq.(VI.21) and (VI.10), we have

$$\sigma_{out} = \text{Sgn}(\gamma - 2) \quad . \tag{VI.22}$$

Thus, we have to consider two possibilities :

1. $1 < \gamma < 2$, i.e. $(\sigma_{in}, \sigma_{out}) = (+1, -1)$

2. $\gamma > 2$, i.e. $(\sigma_{in}, \sigma_{out}) = (+1, +1)$

which correspond to the spacetime conformal diagrams shown in fig.4 and fig.5, respectively.

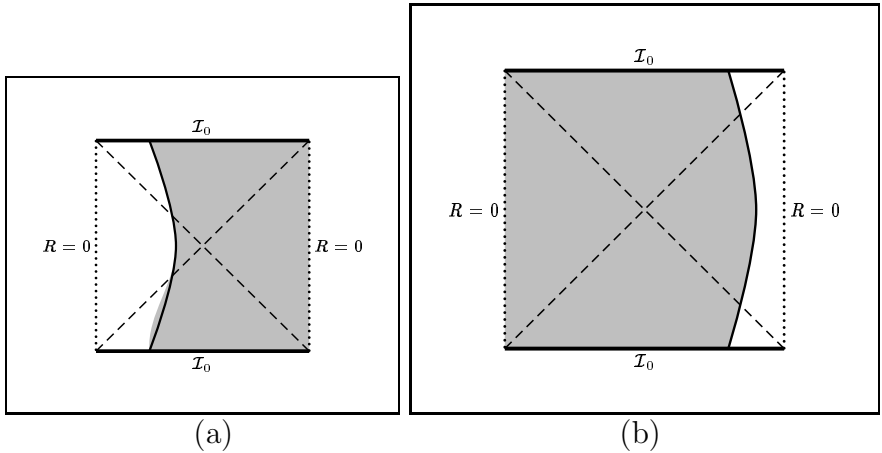

FIG. 4. Penrose diagrams corresponding to the "in" (a), and the "out" (b) domains in the case $1 < \gamma < 2$. The light regions correspond to the physical domains. The side of the diagram in which the trajectories are drawn is determined by the sign of $\sigma_{in}$ and $\sigma_{out}$. Their sign is also related to the direction of the normal to the shell orbit.

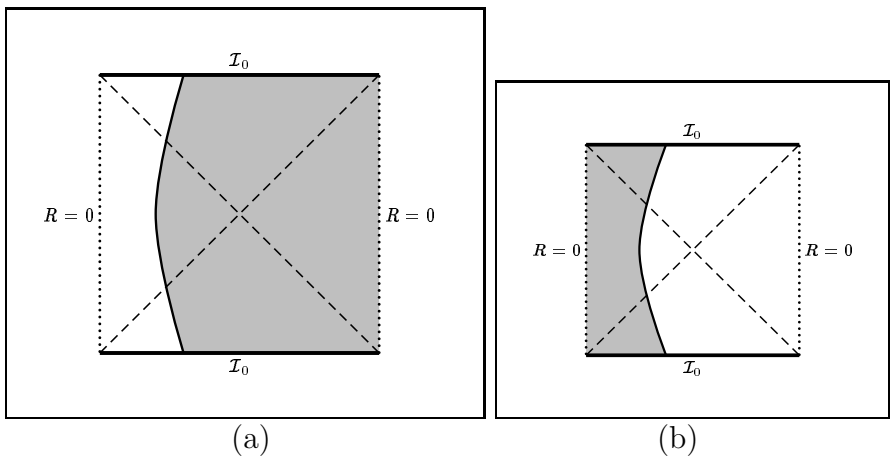

FIG. 5. Penrose diagrams corresponding to the "in" (a), and the "out" (b) domains in the case $\gamma > 2$. Comparing with Fig.(4), note that the change of sign of $\sigma_{out}$ is reflected by the fact that the shell trajectory is now drawn in the opposite half of the conformal diagram.

The physical spacetime is obtained by "gluing" along $\Sigma$ the un-dashed regions of each figure.

Let us consider first case # 2, $\gamma > 2$. From the Hamiltonian of eq.(VI.8), we see that the configuration $R = 0$ (i.e., no shell, the whole spacetime is de Sitter "*out*") satisfies the constraint $H = 0$. On the other hand, our previous analysis of the classical motion revealed that classically allowed configurations for the shell exist only for shell radii $R \geq R_N \equiv \xi_N/\kappa$, where $R_N$ represents the turning point.

Let us recall that false vacuum decay is a quantum process which proceeds via the nucleation of true vacuum bubbles of a given non-vanishing radius which then expand filling the surrounding false vacuum region. In our approach, this process (the formation of one bubble) is described by the quantum mechanical tunneling through the barrier represented in fig.3, from the $R = 0$ shell configuration to one for which $R = R_N$. In terms of conformal diagrams, the process is visualized in fig.6.

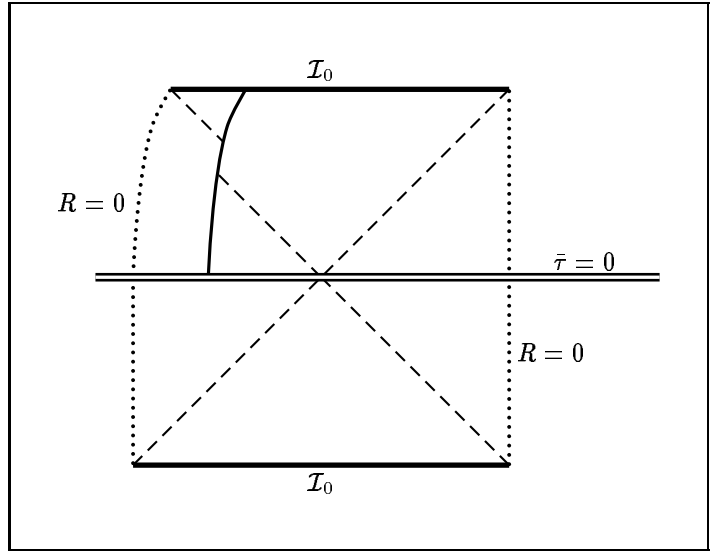

FIG. 6. Penrose diagram corresponding to tunneling along the classically forbidden trajectory of Fig.(3), in the case $\gamma > 2$. Tunneling occurs at $\bar{\tau} = 0$ and nucleates a true vacuum bubble in the false vacuum background.

The probability of this process is given, in the WKB approximation, and apart from an overall factor, by $\exp(-B/\hbar)$, where $B$ is given by eq.(VI.5). In order to calculate the nucleation coefficient $B$ for the present case, it is useful to define

$$y_0^{-1} = -\frac{2}{\alpha + \gamma^2}\left(\gamma - \gamma^2 - \frac{\alpha}{2}\right) \tag{VI.23}$$

$$y_1^{-1} = \frac{\alpha}{\alpha + \gamma^2} \tag{VI.24}$$

$$y_2^{-1} = \frac{4}{\alpha + \gamma^2}\left(\gamma - 1 + \frac{\alpha}{4}\right) \quad . \tag{VI.25}$$

Note that, since $\gamma > 2$, all three constants are positive. The nucleation coefficient $B$ then becomes

$$B = \frac{\xi_N^2}{2G_N}\frac{y_1 y_2}{y_0}\int_0^1 dy \sqrt{y}\sqrt{1-y}\,(y_0 - y)(y_1 - y)^{-1}(y_2 - y)^{-1} \quad . \tag{VI.26}$$

Performing the integration in the complex plane, as described in Appendix C, we obtain

$$B = \frac{\pi \xi_N^2}{2G_N}\frac{y_1 y_2}{y_0}\left\{\left(\frac{y_1}{y_1 - 1}\right)^{\frac{1}{2}}(y_1 - y_0)(y_1 - y_2)^{-1} + \right.$$

$$\left.\left(\frac{y_2}{y_2 - 1}\right)^{\frac{1}{2}}(y_2 - y_0)(y_2 - y_1)^{-1} - 1\right\} \tag{VI.27}$$

$$= \frac{\pi \xi_N^3}{2G_N}\frac{y_1 y_2}{y_0}\left[-\frac{2(1-\gamma)\sqrt{\alpha + \gamma^2}}{2\gamma^2 - 2\gamma + \alpha} - 1\right] \quad . \tag{VI.28}$$

Then, with some lengthy algebra, which we confine to Appendix D, one can show that the nucleation radius and the nucleation coefficient $B$ coincide exactly with the expression obtained by Parke [17].

A simpler case worth considering here, is the limiting value of $B$, given by eq.(VI.27), as $\Lambda_{in} \to 0$. In this case

$$B\left(\Lambda_{in}=0\right) = \frac{\pi \xi_N^3}{2G_N}\frac{y_2}{y_0}\left[\frac{1}{2} + (y_2 - y_0)\left(1 - \left(\frac{y_2}{y_2-1}\right)^{\frac{1}{2}}\right)\right] \quad , \tag{VI.29}$$

which, after unfolding the expression of the constants, takes the form

$$B\left(\Lambda_{in}=0\right) = \frac{\pi^2 \rho}{2} R_N^2 R_0 \quad , \tag{VI.30}$$

where $R_N$ represents the nucleation radius of the bubble,

$$R_N\left(\Lambda_{in}=0\right) = \frac{R_0}{1 + \frac{R_0^2 \Lambda}{12}} \quad . \tag{VI.31}$$

Equations(VI.30-VI.31) reproduce exactly the results obtained by Coleman and de Luccia [16].

As a final remark, and in preparation of the next case, note that in the previous discussion the initial configuration $R = 0$ consists of the classical spacetime obtained by letting $R(\tau) \to 0$ in fig.5.b, with $\sigma_{out} = +1$ during the whole process. Therefore, in that limit the shaded region of fig.5.b disappears, leaving as initial configuration for the tunneling process the whole de Sitter spacetime.

Consider now case # 1, i.e., $\gamma < 2$. In the same limit as before, but keeping now $\sigma_{out} = -1$, we see from fig.4.b that all the un-dashed region (which represents the physical region) disappears. Therefore, in this case we have a peculiar situation in which the initial configuration corresponds to *no* spacetime at all, so that the tunneling process describes now something rather different from vacuum decay, which we interpret as the *creation from nothing* of a closed universe formed by two de Sitter cups.

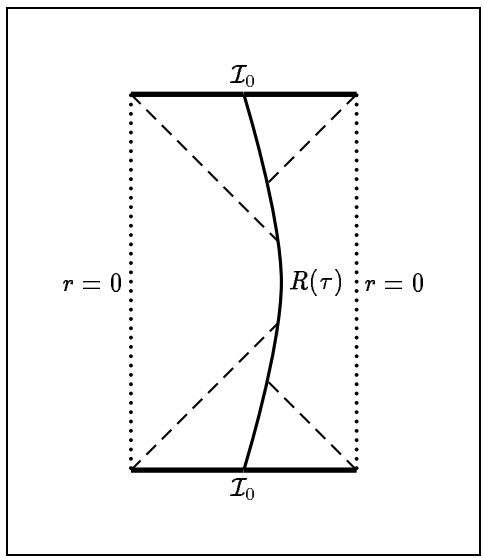

FIG. 7. Penrose diagram corresponding to tunneling along the classically forbidden trajectory of Fig.(3), in the case $1 < \gamma < 2$. In this case the configuration $R \equiv 0$ does not represent a classical solution, and one cannot speak of "bubble nucleation". One possible quantum mechanical interpretation of this process is simply as the birth "from nothing" of a universe composed of two de Sitter domains.

It seems conceivable that this kind of process is characteristic of the foam structure of spacetime at small scales of length at which black holes, wormholes and inflationary domains are continuously created out of the vacuum. The expression for the nucleation coefficient in this case ($\gamma < 2$), is

$$B = \frac{\pi \xi_N^3}{2G_N} \frac{y_1 y_2}{y_0} \left[ -\frac{\sqrt{\alpha + \gamma^2}}{2\gamma - 2\gamma^2 - \alpha} (2\gamma - 2 + \alpha) - 1 \right] \quad . \tag{VI.32}$$

For $\Lambda_{in} = 0$, the above expression reduces to

$$B(\Lambda_{in} = 0) = \frac{\pi \xi_N^3}{2G_N} \frac{\gamma}{4}(\gamma + 1) \quad . \tag{VI.33}$$

By considering a further limit, $\Lambda_{in} = \Lambda_{out} = 0$, corresponding to the *creation* of a *Minkowski pair*, we obtain

$$B_{MM} = \frac{\pi \xi_N^3}{2} \left( \frac{1}{2} - \frac{3}{8y_0} \right) = \frac{\pi}{2} R_0^2 = \frac{1}{8\pi \rho^2 G_N^3} \quad , \tag{VI.34}$$

in exact agreement with the bounce calculation of ref. [18].

## VII. CONCLUSION

In this paper we have formulated the classical and quantum dynamics of a spherically symmetric shell in a de Sitter–Schwarzschild background, directly in terms of the Einstein–Hilbert action supplemented by an arbitrary equation of state. An effective action for the shell radial degree of freedom was then obtained following the FGG–reduction technique. A distinctive feature of our formulation is its invariance under a general redefinition of the evolution parameter. This feature leads to the Hamiltonian constraint $H = 0$. This constraint is discussed further in Appendix A, where we have compared our variational procedure with the FGG–approach, in order to check the consistency of our result. In our formulation, however, the vanishing of $H$, far from being a fortuitous coincidence which makes the *true* classical dynamics and the *naive* classical dynamics identical [19], acquires a precise mathematical and physical significance, namely, that *reparametrization invariance implies that there is no energy associated with the evolution parameter $\tau$*. This, in a nutshell, is the simple message on which our approach is based: once unfolded and reinterpreted by the machinery of the canonical formalism, that message translates into the matching condition(I.1), and effectively controls the classical evolution of the system which was briefly reviewed in Section IV. At the quantum level, however, any two approaches based on a different choice of evolution parameter may differ significantly. Thus, building on our classical result, we have laid the foundations of our quantum approach and explored some of its consequences. We have shown that the vanishing of the Hamiltonian, in a weak sense, can now be interpreted as the Wheeler-De Witt equation $\hat{H}|\Psi\rangle = 0$ for the physical states. However, the explicit construction of $\hat{H}$ as an hermitian operator acting on a Hilbert space is by no means straightforward. The sign multiplicity of the $\beta$–functions, non-locality and ordering ambiguities are the major limitations to the full utilization of our formulation. However, to the extent that our approach is analogous to the minisuperspace approach to quantum cosmology, it seems to us that the above difficulties may represent a shadow of deeper problems which are widely suspected to be a general feature of quantum gravity [15]. At any rate, because of the above difficulties, we limited our considerations to the quantum dynamics of a shell in the WKB approximation. In particular, we have studied the probability of quantum tunneling under the classical potential barrier, and have shown that vacuum decay can be described by such a tunneling process. All known results on decay probabilities and nucleation radii are correctly reproduced by our formalism. In addition, we have speculated on the nature of some rather exotic processes which we have interpreted as "creation of vacuum domains from nothing". However, a proper treatement of such processes lies beyond our first quantized formulation of shell dynamics. One possible step in this direction would be to apply the Dirac formalism of canonical quantization not only to the shell, but to the gravitational field as well [14,15], [20]. In such a case, the whole spacetime enters the theory as a geometrodynamical entity.

After this paper was submitted for publication, a similar formulation was proposed in [21] to describe massive dust shells.

## APPENDIX A: THE FGG–METHOD AND THE HAMILTONIAN CONSTRAINT

In this appendix we wish to discuss the relationship between our variational procedure and the FGG–approach. In this connection, it seems worth emphasizing that our equations of motion were obtained by extremizing the action $S_{eff}$ with respect to arbitrary variations of the functions $N(\tau)$ and $R(\tau)$, vanishing at the end points, while the temporal boundary was held fixed. This variational procedure is consistent with our minisuperspace approach to the intrinsic dynamics of the shell, according to which there should be no reference to the internal, or external time coordinate, or to the particular volume of integration chosen, since the latter is just a convenient device to arrive at the equations of motion for the shell. This represents a departure from the FGG–method [7], which is based on a different variational procedure. In that approach, the evolution of the shell is parametrized by the external time, so that it seems natural to consider variations in which the initial and final external time $t_i$ and $t_f$ are held fixed. This, however, means that $\tau_f$ is not fixed, but varies according to

$$\delta\tau_f = \frac{\int_{\tau_i}^{\tau_f} \left\{ \left[\frac{\partial F}{\partial R} - \frac{d}{d\tau}\left(\frac{\partial F}{\partial \dot{R}}\right)\right]\delta R(\tau) + \frac{\partial F}{\partial N}\delta N(\tau) \right\} d\tau}{F(\tau_f) - \left(\dot{R}\frac{\partial F}{\partial \dot{R}}\right)_f} \quad . \tag{A.1}$$

This equation is obtained by taking the variation of eq.(II.10)

$$t_f - t_i = \int_{\tau_i}^{\tau_f} F\left(R, \dot{R}, N\right) d\tau \quad , \tag{A.2}$$

where

$$F \equiv \frac{\beta_{out}}{A_{out}} = \frac{\sqrt{N^2 A_{out} + \dot{R}^2}}{A_{out}} \quad. \tag{A.3}$$

Since now $\tau_f$ is not fixed, the variation of $S_{eff}$ yelds additional contributions

$$\delta S = \int_{\tau_i}^{\tau_f} \left\{ \left[ \frac{\partial L}{\partial R} - \frac{d}{d\tau}\left(\frac{\partial L}{\partial \dot{R}}\right) \right] \delta R(\tau) + \frac{\partial L}{\partial N} \delta N(\tau) \right\} d\tau +$$
$$+ L(\tau_f)\delta\tau_f + \left(\frac{\partial L}{\partial \dot{R}}\right)_f \delta R_f \quad, \tag{A.4}$$

where

$$\delta R_f = -\dot{R}(\tau_f)\delta\tau_f \quad. \tag{A.5}$$

Inserting eqs.(A.1) and (A.5) into eq.(A.4), we obtain

$$\delta S = \int_{\tau_i}^{\tau_f} \left\{ \left[ \frac{\partial L}{\partial R} - \frac{d}{d\tau}\left(\frac{\partial L}{\partial \dot{R}}\right) - \frac{H_f}{\xi_f}\frac{\partial F}{\partial R} - \frac{d}{d\tau}\left(\frac{\partial f}{\partial \dot{R}}\right) \right] \delta R(\tau) + $$
$$+ \left( \frac{\partial L}{\partial N} - \frac{H_f}{\xi_f}\frac{\partial F}{\partial N} \right) \delta N(\tau) \right\} d\tau \quad, \tag{A.6}$$

where

$$\xi_f = \left(F - \dot{R}\frac{\partial F}{\partial \dot{R}}\right)_f \tag{A.7}$$

and $H_f \equiv H(\tau_f)$. Demanding that $\delta S$ vanishes, results now in the following equations of motion

$$\frac{\partial L}{\partial N} - \frac{H_f}{\xi_f}\frac{\partial F}{\partial N} = 0 \tag{A.8}$$

$$\frac{\partial L}{\partial R} - \frac{d}{d\tau}\left(\frac{\partial L}{\partial \dot{R}}\right) - \frac{H_f}{\xi_f}\frac{\dot{R}}{N}\left[\frac{\partial F}{\partial R} - \frac{d}{d\tau}\left(\frac{\partial F}{\partial \dot{R}}\right)\right] = 0 \quad. \tag{A.9}$$

Now, Eq.(A.8) does not include an acceleration term; again, it represents a constraint which can be rewritten as

$$\frac{H}{N} + \frac{H_f}{\xi_f}\frac{\partial F}{\partial N} = 0 \quad. \tag{A.10}$$

Eq.(A.9), on the other hand, gives

$$\frac{d}{d\tau}\left(\frac{H}{N}\right) - \frac{H_f}{\xi_f}\frac{\dot{R}}{N}\left[\frac{\partial F}{\partial R} - \frac{d}{d\tau}\left(\frac{\partial F}{\partial \dot{R}}\right)\right] = 0 \quad, \tag{A.11}$$

so that the solution of both equations (A.10-A.11) is $H = 0$. In fact, setting $H = 0$ in eq.(A.10) we find $H_f = 0$, which, once inserted in eq.(A.11), implies that the Hamiltonian constraint $H = 0$ is preserved in time, which is our result.

# APPENDIX B: HAMILTONIAN FOR $A_{IN} \gtrless 0$, $A_{OUT} \gtrless 0$

For the sake of notational simplicity, let us set $G_N = 1$. We start with the expression of the momentum in the case $A_{in} > 0$ and $A_{out} > 0$:

$$
\begin{aligned}
\frac{P_R}{R} &= \tanh^{-1}\left(\frac{\dot{R}(\sigma_{in}\beta_{in} - \sigma_{out}\beta_{out})}{\sigma_{in}\sigma_{out}\beta_{in}\beta_{out} - \dot{R}^2}\right) \\
&= \frac{1}{2}\ln\left[\frac{(\sigma_{in}\beta_{in} - \dot{R})(\sigma_{out}\beta_{out} + \dot{R})}{(\sigma_{in}\beta_{in} + \dot{R})(\sigma_{out}\beta_{out} - \dot{R})}\right] \\
&= \ln\left[\frac{(\sigma_{in}\beta_{in} - \dot{R})^2(\sigma_{out}\beta_{out}^2 - \dot{R}^2)}{(\sigma_{in}\beta_{in}^2 - \dot{R}^2)(\sigma_{out}\beta_{out} + \dot{R})^2}\right]^{\frac{1}{2}} \\
&= \ln\left[\sqrt{\frac{A_{out}}{A_{in}}} \frac{|\sigma_{in}\beta_{in} - \dot{R}|}{|\sigma_{out}\beta_{out} - \dot{R}|}\right] \\
^1 &= \ln\left[\sqrt{\frac{A_{out}}{A_{in}}} \frac{(\sigma_{in}\beta_{in} - \dot{R})\sigma_{in}}{(\sigma_{out}\beta_{out} - \dot{R})\sigma_{out}}\right] \\
&= \ln\left[\sigma_{in}\sigma_{out}\sqrt{\frac{A_{out}}{A_{in}}} \frac{\sigma_{in}\beta_{in} - \dot{R}}{\sigma_{out}\beta_{out} - \dot{R}}\right] \quad .
\end{aligned}
\tag{B.1}
$$

Then, enlisting the equalities

$$
e^{\frac{P_R}{R}} - \sigma_{in}\sigma_{out}\sqrt{\frac{A_{out}}{A_{in}}} = \sigma_{in}\sigma_{out}\sqrt{\frac{A_{out}}{A_{in}}}\left(\frac{\sigma_{in}\beta_{in} - \sigma_{out}\beta_{out}}{\sigma_{out}\beta_{out} - \dot{R}}\right) \tag{B.2}
$$

$$
e^{-\frac{P_R}{R}} - \sigma_{in}\sigma_{out}\sqrt{\frac{A_{out}}{A_{in}}} = \sigma_{in}\sigma_{out}\sqrt{\frac{A_{out}}{A_{in}}}\left(\frac{\sigma_{in}\beta_{in} - \sigma_{out}\beta_{out}}{\sigma_{out}\beta_{out} + \dot{R}}\right) \quad , \tag{B.3}
$$

so that

$$
\begin{aligned}
1 + \frac{A_{out}}{A_{in}} - 2\sigma_{in}\sigma_{out}\sqrt{\frac{A_{out}}{A_{in}}}\cosh\left(\frac{P_R}{R}\right) &= \\
= \left(e^{\frac{P_R}{R}} - \sigma_{in}\sigma_{out}\sqrt{\frac{A_{out}}{A_{in}}}\right)\left(e^{-\frac{P_R}{R}} - \sigma_{in}\sigma_{out}\sqrt{\frac{A_{out}}{A_{in}}}\right) &= \\
&= \frac{1}{A_{in}}\left(\frac{H - \kappa R^2}{R}\right)^2 \quad ,
\end{aligned}
\tag{B.4, B.5}
$$

---

[1] This particulaer step is based on the following observation: $\sigma_{in}\beta_{in} - \dot{R} = \sigma_{in}\sqrt{\dot{R}^2 + A_{in}} - \dot{R}$ with $A_{in} > 0$, so that the argument of the absolute value is positive if $\sigma_{in} > 0$, and negative if $\sigma_{in} < 0$, i.e., it has the same sign of $\sigma_{in}\beta_{in}$, that is $\sigma_{in}$. The same is true for the denominator.

we find the explicit form of the Hamiltonian quoted in the text (Eq.IV.12):

$$H = \kappa R^2 - R\left[A_{in} + A_{out} - 2\sigma_{in}\sigma_{out}\sqrt{A_{in}A_{out}}\cosh\left(\frac{P_R}{R}\right)\right]^{\frac{1}{2}} \quad . \tag{B.6}$$

We now repeat the same steps for the case $A_{in} > 0$ and $A_{out} < 0$. Letting $\bar{A}_{out} = -A_{out}$, we find

$$\begin{aligned}\frac{P_R}{R} &= \tanh^{-1}\left(\frac{\sigma_{in}\sigma_{out}\beta_{in}\beta_{out} - \dot{R}^2}{\dot{R}(\sigma_{in}\beta_{in} - \sigma_{out}\beta_{out})}\right) \\ &= \frac{1}{2}\ln\left[\frac{(\sigma_{in}\beta_{in} - \dot{R})(\sigma_{out}\beta_{out} + \dot{R})}{(\dot{R} + \sigma_{in}\beta_{in})(\dot{R} - \sigma_{out}\beta_{out})}\right] \\ &= \ln\left[\frac{(\sigma_{in}\beta_{in} - \dot{R})^2(\sigma_{out}\beta_{out}^2 - \dot{R}^2)}{(\sigma_{in}\beta_{in}^2 - \dot{R}^2)(\sigma_{out}\beta_{out} - \dot{R})^2}(-)\right]^{\frac{1}{2}} \\ &= \ln\left[\sqrt{\frac{\bar{A}_{out}}{A_{in}}}\frac{|\sigma_{in}\beta_{in} - \dot{R}|}{|\sigma_{out}\beta_{out} - \dot{R}|}\right] \\ ^2 &= \ln\left[\sqrt{\frac{\bar{A}_{out}}{A_{in}}}\frac{(\sigma_{in}\beta_{in} - \dot{R})\sigma_{in}}{(\sigma_{out}\beta_{out} - \dot{R})(-)}\right] \\ &= \ln\left[-\sigma_{in}\sqrt{\frac{\bar{A}_{out}}{A_{in}}}\frac{\sigma_{in}\beta_{in} - \dot{R}}{\sigma_{out}\beta_{out} - \dot{R}}\right] \quad . \end{aligned} \tag{B.7}$$

Enlisting now the equalities

$$e^{\frac{P_R}{R}} + \sigma_{in}\sqrt{\frac{\bar{A}_{out}}{A_{in}}} = \sigma_{in}\sqrt{\frac{\bar{A}_{out}}{A_{in}}}\left(\frac{\sigma_{in}\beta_{in} - \sigma_{out}\beta_{out}}{\sigma_{out}\beta_{out} - \dot{R}}\right)(-) \tag{B.8}$$

$$e^{-\frac{P_R}{R}} - \sigma_{in}\sqrt{\frac{\bar{A}_{out}}{A_{in}}} = \sigma_{in}\sqrt{\frac{\bar{A}_{out}}{A_{in}}}\left(\frac{\sigma_{in}\beta_{in} - \sigma_{out}\beta_{out}}{\sigma_{out}\beta_{out} + \dot{R}}\right) \quad , \tag{B.9}$$

so that

$$1 - \frac{\bar{A}_{out}}{A_{in}} - 2\sigma_{in}\sqrt{\frac{\bar{A}_{out}}{A_{in}}}\sinh\left(\frac{P_R}{R}\right) =$$

---

[2]We work with the numerator as described in the previous footnote. For the denominator we follow the same procedure, but due to the fact that now $A_{out} < 0$, the sign of $\dot{R}$ dominates.

$$= \left(e^{\frac{P_R}{R}} + \sigma_{in}\sqrt{\frac{\bar{A}_{out}}{A_{in}}}\right)\left(e^{-\frac{P_R}{R}} - \sigma_{in}\sqrt{\frac{\bar{A}_{out}}{A_{in}}}\right) = \tag{B.10}$$

$$= \frac{1}{A_{in}}\left(\frac{H - (4\pi\rho)R^2}{R}\right)^2 \quad, \tag{B.11}$$

we find

$$H = \kappa R^2 - R\left[A_{in} + A_{out} - 2\sigma_{in}\sqrt{-A_{in}A_{out}}\sinh\left(\frac{P_R}{R}\right)\right]^{\frac{1}{2}}. \tag{B.12}$$

Evidently, we can follow the same procedure in the cases $A_{in} < 0$, $A_{out} < 0$; $A_{in} < 0$, $A_{out} > 0$. The corresponding expressions are

$$H = \kappa R^2 - R\left[A_{in} + A_{out} - 2\sqrt{A_{in}A_{out}}\cosh\left(\frac{P_R}{R}\right)\right]^{\frac{1}{2}} \tag{B.13}$$
$$\text{if}\quad A_{in} < 0,\ A_{out} < 0$$

$$H = \kappa R^2 - R\left[A_{in} + A_{out} - 2\sigma_{out}\sqrt{-A_{in}A_{out}}\sinh\left(\frac{P_R}{R}\right)\right]^{\frac{1}{2}} \tag{B.14}$$
$$\text{if}\quad A_{in} < 0,\ A_{out} > 0\quad,$$

or, in a more compact notation

$$H = \kappa R^2 - \mathrm{Sgn}(\rho)R\cdot$$
$$\cdot\left[A_{in} + A_{out} - 2\sigma_{in}\sigma_{out}\left(\frac{|A_{in}A_{out}|}{1 - \left[\tanh\left(\frac{P_R}{R}\right)\right]^{2\frac{A_{in}A_{out}}{|A_{in}A_{out}|}}}\right)^{\frac{1}{2}}\right]^{\frac{1}{2}}. \tag{B.15}$$

The Hamiltonian for the shell of dust quoted in Section IV can be derived along the same steps with little change, and one obtains

$$H = m - \mathrm{Sgn}(m)R\cdot$$
$$\cdot\left[A_{in} + A_{out} - 2\sigma_{in}\sigma_{out}\left(\frac{|A_{in}A_{out}|}{1 - \left[\tanh\left(\frac{P_R}{R}\right)\right]^{2\frac{A_{in}A_{out}}{|A_{in}A_{out}|}}}\right)^{\frac{1}{2}}\right]^{\frac{1}{2}}. \tag{B.16}$$

**APPENDIX C: BASIC INTEGRAL FOR THE DE SITTER–DE SITTER CASE.**

The calculation of the integral (VI.26) is performed in the complex plane by considering the function

$$f(z) = \sqrt{z}\sqrt{z-1}\,(z-y_0)\,(z-y_1)^{-1}\,(z-y_2)^{-1} \tag{C.1}$$

which has two branch points ($z = 0$ and $z = 1$), and two simple poles ($z = y_1$ and $z = y_2$). Accordingly, we integrate along the path $\Gamma_{\epsilon_1,\epsilon_2}$:

$C_{\epsilon_1}(0)$ : a clockwise circumference of radius $\epsilon_1$ and center $z = 0$;

$0 + \epsilon_1 \to 1 - \epsilon_2$: an oriented line segment of the x-axis in the upper half of the complex plane;

$C_{\epsilon_2}(1)$: a clockwise circumference of radius $\epsilon_2$ and center $z = 1$;

$0 + \epsilon_1 \leftarrow 1 - \epsilon_2$: an oriented line segment of the x-axis in the lower half of the complex plane

with

$$0 \leq \arg(z) < 2\pi \tag{C.2}$$

$$0 \leq \arg(z-1) < 2\pi \quad . \tag{C.3}$$

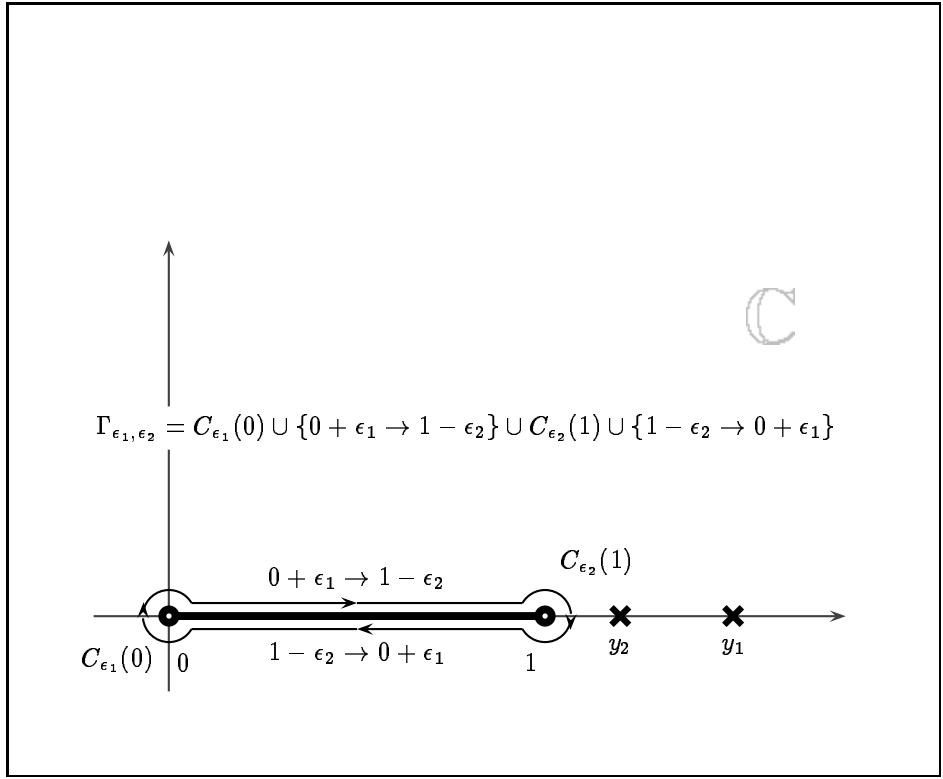

FIG. 8. Integration path $\Gamma_{\epsilon_1,\epsilon_2}$

Then, we have

$$\lim_{\epsilon_1,\epsilon_2\to 0}\int_{\Gamma_{\epsilon_1,\epsilon_2}} dz f(z) = 2i\pi \left[\mathcal{R}\text{es}\{f,+\infty\} + \mathcal{R}\text{es}\{f,y_1\} + \mathcal{R}\text{es}\{f,y_2\}\right] \quad . \tag{C.4}$$

It is also true that

$$\lim_{\epsilon_1,\epsilon_2\to 0}\int_{\Gamma_{\epsilon_1,\epsilon_2}} dz f(z) = \int_0^1 dz f_+(z) - \int_0^1 dz f_-(z) \tag{C.5}$$

with

$$f_+(z) = [|z|]^{\frac{1}{2}}\left[|z-1|e^{i\pi}\right]^{-\frac{1}{2}}\left[|z-y_0|e^{i\pi}\right]\left[|z-y_1|e^{i\pi}\right]^{-1}\left[|z-y_2|e^{i\pi}\right]^{-1}$$
$$= if(y) \tag{C.6}$$

$$f_-(z) = \left[|z|e^{2i\pi}\right]^{\frac{1}{2}}\left[|z-1|e^{i\pi}\right]^{-\frac{1}{2}}\left[|z-y_0|e^{i\pi}\right]\left[|z-y_1|e^{i\pi}\right]^{-1}\left[|z-y_2|e^{i\pi}\right]^{-1}$$
$$= -if(y) \quad . \tag{C.7}$$

Therefore,

$$\lim_{\epsilon_1,\epsilon_2\to 0}\int_{\Gamma_{\epsilon_1,\epsilon_2}} dz f(z) = 2iI \tag{C.8}$$

with

$$I = \pi\left[\mathcal{R}\text{es}\{f,+\infty\} + \mathcal{R}\text{es}\{f,y_1\} + \mathcal{R}\text{es}\{f,y_2\}\right] \quad . \tag{C.9}$$

The residue of the function at infinity is found from its asymptotic expansion in powers of $z^{-1}$:

$$f(z) = z^{\frac{1}{2}}\frac{1}{(z-1)^{\frac{1}{2}}}(z-y_0)\frac{1}{z-y_1}\frac{1}{z-y_2}$$
$$= z^{\frac{1}{2}}\frac{1}{z^{\frac{1}{2}}\left(1-\frac{1}{z}\right)^{\frac{1}{2}}}z\left(1-\frac{y_0}{z}\right)\frac{1}{z\left(1-\frac{y_1}{z}\right)}\frac{1}{z\left(1-\frac{y_1}{z}\right)}$$
$$\sim \frac{1}{z}\left(1+\frac{1}{2z}+\frac{3}{8z^2}+\mathcal{O}\left(z^{-3}\right)\right) \cdot$$
$$\cdot \left(1+\frac{y_1}{z}+\frac{y_1^2}{z^2}+\mathcal{O}\left(z^{-3}\right)\right)\left(1+\frac{y_2}{z}+\frac{y_2^2}{z^2}+\mathcal{O}\left(z^{-3}\right)\right)$$
$$\sim \frac{1}{z}+\mathcal{O}\left(\frac{1}{z^2}\right) \tag{C.10}$$

The opposite of the coefficient of $z^{-1}$ yields the result:

$$\mathcal{R}\text{es}\{f,+\infty\} = -1 \quad . \tag{C.11}$$

Next, the residues at the poles $z = y_1, y_2$ are given by

$$\mathcal{R}\text{es}\left\{f, y_1\right\} = y_1^{\frac{1}{2}} (y_1 - 1)^{-\frac{1}{2}} (y_1 - y_0)(y_1 - y_2)^{-1} \tag{C.12}$$

$$\mathcal{R}\text{es}\left\{f, y_2\right\} = y_2^{\frac{1}{2}} (y_2 - 1)^{-\frac{1}{2}} (y_2 - y_0)(y_2 - y_1)^{-1} \quad . \tag{C.13}$$

Then, summing up our results, we finally obtain the expression of the integral

$$I = \pi \left\{ \left(\frac{y_1}{y_1 - 1}\right)^{\frac{1}{2}} (y_1 - y_0)(y_1 - y_2)^{-1} + \right.$$
$$\left. + \left(\frac{y_2}{y_2 - 1}\right)^{\frac{1}{2}} (y_2 - y_0)(y_2 - y_1)^{-1} - 1 \right\} \quad . \tag{C.14}$$

### APPENDIX D: COMPARISON WITH PARKE

Parke's result for the bounce action is [17]

$$B = B_0 \, r\left[\left(\frac{\bar{\rho}_0}{2\Lambda}\right)^2, \frac{\Lambda^2}{\lambda^2}\right] \quad , \tag{D.1}$$

where

$$r\left[x, y\right] = \frac{2\left[(1 + xy) - (1 + 2xy + x^2)^{\frac{1}{2}}\right]}{x^2 \left(y^2 - 1\right)(1 + 2xy + x^2)^{\frac{1}{2}}} \quad . \tag{D.2}$$

Moreover

$$\bar{\rho}_0 = \frac{3S_1}{U_f - U_t} \tag{D.3}$$

is the critical radius in the absence of gravity, and

$$U_{f/t} = U\left(\phi_{f/t}\right) \tag{D.4}$$

represent the extremal points of the potential $U(\phi)$ for the scalar field $\phi$, which is minimally coupled to gravity; they correspond to the *false* and *true vacuum* respectively. The quantities $\Lambda$ e $\lambda$ are related to these parameters by the following relations[3]

$$\Lambda^2 = \left[\frac{8\pi \left(U_f - U_t\right)}{3}\right]^{-1} \tag{D.5}$$

$$\lambda^2 = \left[\frac{8\pi \left(U_f + U_t\right)}{3}\right]^{-1} \tag{D.6}$$

---

[3]We set $G \equiv \hbar \equiv 1$

and

$$B_0 = \frac{27\pi^2 S_1^4}{2(U_f - U_t)^3} \tag{D.7}$$

represents the bounce action in the absence of gravity.

On the other hand, our result is

$$B = \frac{\pi \xi_N^3}{2} \frac{y_1 y_2}{y_0} \left\{ \left(\frac{y_1}{y_1 - 1}\right)^{\frac{1}{2}} (y_1 - y_0)(y_1 - y_2)^{-1} + \right.$$
$$\left. + \left(\frac{y_2}{y_2 - 1}\right)^{\frac{1}{2}} (y_2 - y_0)(y_2 - y_1)^{-1} - 1 \right\} \quad , \tag{D.8}$$

in terms of the parameters defined in eqs.(VI.9-VI.10)-(VI.18)-(VI.23- VI.25). The correspondence between our cosmological constants and the *false vacuum/true vacuum* energies is given by the relations:

$$\epsilon_{in} = U_f \quad , \tag{D.9}$$
$$\epsilon_{out} = U_t \quad , \tag{D.10}$$

with

$$\Lambda_{in} = 8\pi \epsilon_{in} \quad , \tag{D.11}$$
$$\Lambda_{out} = 8\pi \epsilon_{out} \quad . \tag{D.12}$$

Furthermore,

$$R_0 = \frac{3|\rho|}{|\Delta \epsilon|} = \frac{2}{|4\pi\rho|} \frac{1}{|1 - \gamma|} \tag{D.13}$$

corresponds to $\bar{\rho}_0$ which, in turn, enables us to identify:

$$S_1 \to \rho \tag{D.14}$$
$$\Delta \epsilon \to U_f - U_t \quad . \tag{D.15}$$

Using this translation code, one can verify that Parke's expression for the nucleation coefficient in the absence of gravity corresponds to

$$B_0 = \frac{27\pi^2 S_1^4}{2(U_f - U_t)^3} \quad \to \quad \frac{\pi^2}{2} \left(\frac{3|\rho|}{\Delta \epsilon}\right)^3 \frac{\rho^4}{|\rho|^3} = \frac{\pi^2}{2} R_0^3 |\rho| \quad , \tag{D.16}$$

which is exactly our result. Passing to the more general case, we note that:

$$\Lambda^2 \to \left(\frac{\Lambda_{in} - \Lambda_{out}}{3}\right)^{-1} = \frac{1}{\kappa^2 (1 - \gamma)} \tag{D.17}$$

$$\lambda^2 \to \left(\frac{\Lambda_{in} + \Lambda_{out}}{3}\right)^{-1} = \frac{2}{\kappa^2 (2\gamma - 2 + \alpha)} \tag{D.18}$$

from which we deduce,

$$\frac{\Lambda^2}{\lambda^2} \to \frac{2\gamma - 2 + \alpha}{2(1-\gamma)} \tag{D.19}$$

$$\frac{\bar{\rho}_0^2}{4\Lambda^2} \to \frac{R_0^2}{4}\kappa^2(1-\gamma) \tag{D.20}$$

and, as a consequence,

$$x \to \frac{1}{1-\gamma} \tag{D.21}$$

$$y \to \frac{2\gamma - 2 + \alpha}{2(1-\gamma)} \tag{D.22}$$

$$x^2 \to \frac{1}{(1-\gamma)^2} \tag{D.23}$$

$$xy \to \frac{2\gamma - 2 + \alpha}{2(1-\gamma)^2} \tag{D.24}$$

$$1 + xy \to \frac{2\gamma^2 - 2\gamma + \alpha}{2(1-\gamma)^2} = \frac{\alpha + \gamma^2}{2y_0(1-\gamma)^2} \tag{D.25}$$

$$\sqrt{1 + 2xy + x^2} \to \frac{\sqrt{\alpha + \gamma^2}}{|1-\gamma|} \tag{D.26}$$

$$y^2 - 1 \to \frac{\alpha(4\gamma - 4 + \alpha)}{4(1-\gamma)^2} = \frac{\alpha(\alpha + \gamma^2)}{4y_2(1-\gamma)^2} \quad . \tag{D.27}$$

Substituting all of the above in Eq.(D.1), we find

$$B(\text{Parke}) \to$$

$$\to B_0 \frac{2\left[\frac{\alpha + \gamma^2}{2y_0(1-\gamma)^2} - \frac{\sqrt{\alpha + \gamma^2}}{|1-\gamma|}\right]}{\frac{1}{(1-\gamma)^2}\frac{\alpha(\alpha + \gamma^2)}{4y_2(1-\gamma)^2}\frac{\sqrt{\alpha + \gamma^2}}{|1-\gamma|}} =$$

$$= B_0 \frac{8(1-\gamma)^4 y_2}{\alpha(\alpha + \gamma^2)}\left[-\frac{\sqrt{\alpha + \gamma^2}(2\gamma - 2\gamma^2 - \alpha)}{2|1-\gamma|(\alpha + \gamma^2)} - 1\right] =$$

$$= B_0 \frac{4|1-\gamma|^3(2\gamma - 2\gamma^2 - \alpha) y_2}{\alpha(\alpha + \gamma^2)^{\frac{3}{2}}}\left[-1 - \frac{2|1-\gamma|\sqrt{\alpha + \gamma^2}}{2\gamma^2 - 2\gamma + \alpha}\right] =$$

$$= 4B_0 \frac{y_2 y_1}{y_0}\frac{|1-\gamma|^3}{(\alpha + \gamma^2)^{\frac{3}{2}}}\left[\frac{2|1-\gamma|\sqrt{\alpha + \gamma^2}}{2\gamma - 2\gamma^2 - \alpha} - 1\right] \quad . \tag{D.28}$$

Next, we proceed in the same fashion with our own result. First, we have

$$y_1 - y_0 = \frac{2\gamma(1-\gamma)(\alpha + \gamma^2)}{\alpha(2\gamma - 2\gamma^2 - \alpha)} \tag{D.29}$$

$$y_1 - y_2 = \frac{4(\gamma - 1)(\alpha + \gamma^2)}{\alpha(4\gamma - 4 + \alpha)} \tag{D.30}$$

$$y_2 - y_0 = \frac{2(2-\gamma)(\gamma-1)(\alpha+\gamma^2)}{(4\gamma-4+\alpha)(2\gamma-2\gamma^2-\alpha)} \tag{D.31}$$

$$\frac{y_1}{y_1-1} = \frac{\alpha+\gamma^2}{\gamma^2} \tag{D.32}$$

$$\frac{y_2}{y_2-1} = \frac{\alpha+\gamma^2}{(\gamma-2)^2} \quad. \tag{D.33}$$

Then, in the case in which $\gamma < 0$, or $\gamma > 2$ our expression for the nucleation coefficient becomes

$$B = \frac{\pi\xi_N^3}{2(4\pi\rho)^2} \frac{y_1 y_2}{y_0} \cdot$$
$$\cdot \left\{ \frac{\sqrt{\alpha+\gamma^2}}{|\gamma|} \frac{2\gamma(1-\gamma)(\alpha+\gamma^2)}{\alpha(2\gamma-2\gamma^2-\alpha)} \frac{\alpha(4\gamma-4+\alpha)}{4(\alpha+\gamma^2)(\gamma-1)} + \right.$$
$$- \frac{\sqrt{\alpha+\gamma^2}}{|2-\gamma|} \frac{2(\alpha+\gamma^2)(2-\gamma)(\gamma-1)}{(4\gamma-4+\alpha)(2\gamma-2\gamma^2-\alpha)} \cdot$$
$$\left. \cdot \frac{\alpha(4\gamma-4+\alpha)}{4(\alpha+\gamma^2)(\gamma-1)} - 1 \right\} =$$
$$= \frac{\pi\xi_N^3}{2(4\pi\rho)^2} \frac{y_1 y_2}{y_0} \left[ -\frac{\sqrt{\alpha+\gamma^2}}{2\gamma-2\gamma^2-\alpha} \cdot \right.$$
$$\left. \cdot \left( \frac{\gamma}{|\gamma|}(4\gamma-4+\alpha) + \frac{2-\gamma}{|2-\gamma|}\alpha \right) - 1 \right] =$$
$$= \frac{\pi\xi_N^3}{2(4\pi\rho)^2} \frac{y_1 y_2}{y_0} \left[ \frac{2|1-\gamma|\sqrt{\alpha+\gamma^2}}{2\gamma^2-2\gamma+\alpha} - 1 \right] \quad. \tag{D.34}$$

Finally, comparing the expressions (D.28)–(D.34), one sees that their equivalence is related to the equivalence of the terms

$$\frac{\pi\xi_N^3}{2(4\pi\rho)^2} \quad \text{and} \quad \frac{8|1-\gamma|^3}{2(\alpha+\gamma^2)^{\frac{3}{2}}} B_0 \quad . \tag{D.35}$$

However, in view of the equivalence between eqs.(D.16) and (VI.18), the only equality that needs to be proved is that between

$$\frac{8\pi}{2(4\pi\rho)^2(\alpha+\gamma^2)^{\frac{3}{2}}} \quad \text{and} \quad \frac{8|1-\gamma|^3}{2(\alpha+\gamma^2)^{\frac{3}{2}}} \frac{\pi^2}{2} \frac{8|\rho|}{|4\pi\rho|^3|1-\gamma|^3} \tag{D.36}$$

which is trivially true.

In order to complete our discussion, we note that in the case $0 < \gamma < 2$ our result and that of Parke do not coincide. The reason for this discrepancy may be traced back to the fact that in the given interval of variation of $\gamma$, one finds $\sigma_{in} = -\sigma_{out}$. Therefore, as discussed in the text, the shell configuration $R \equiv 0$ is no longer a classical solution so that no quantum tunneling can occur in the first place.

Figure Captions

FIG. 9. Graphical representation of the integration volume with two dimensions suppressed: $T_i$ ($T_f$) and $t_i$ ($t_f$) label the initial (final) surfaces which are the space–like boundaries of the integration volume. $R_1$ and $R_2$ represent the time–like boundaries, and $R(\tau)$ parametrizes the shell radius which separates the interior domain from the exterior one.

FIG. 10. Graphical representation, in a generic case, of the effective potential described by equation (IV.5). Typically, there are three distinct trajectories. Two of them are classically allowed: (A) bounded, and (C) unbounded. Trajectory (B) is classically forbidden and corresponds to quantum mechanical tunneling through the potential barrier. The exponential of the action evaluated along B gives the WKB approximation to the tunneling amplitude from A to C.

FIG. 11. Graphical representation of the effective potential $V(\xi)$ for the situation in which the "in" and "out" geometries are of de Sitter type. $\xi_N$ will be referred to as the "nucleation radius". The diagram displays only the unbounded trajectory, since the bounded one is degenerate ($R \equiv 0$). Note that only for $\sigma_{in} = \sigma_{out}$ this is a classical solution of eq.(IV.5).

FIG. 12. Penrose diagrams corresponding to the "in" (a), and the "out" (b) domains in the case $1 < \gamma < 2$. The light regions correspond to the physical domains. The side of the diagram in which the trajectories are drawn is determined by the sign of $\sigma_{in}$ and $\sigma_{out}$. Their sign is also related to the direction of the normal to the shell orbit.

FIG. 13. Penrose diagrams corresponding to the "in" (a), and the "out" (b) domains in the case $1 < \gamma < 2$. The light regions correspond to the physical domains. The side of the diagram in which the trajectories are drawn is determined by the sign of $\sigma_{in}$ and $\sigma_{out}$. Their sign is also related to the direction of the normal to the shell orbit.

FIG. 14. Penrose diagrams corresponding to the "in" (a), and the "out" (b) domains in the case $\gamma > 2$. Comparing with Fig.(4), note that the change of sign of $\sigma_{out}$ is reflected by the fact that the shell trajectory is now drawn in the opposite half of the conformal diagram.

FIG. 15. Penrose diagram corresponding to tunneling along the classically forbidden trajectory of Fig.(3), in the case $\gamma > 2$. Tunneling occurs at $\bar{\tau} = 0$ and nucleates a true vacuum bubble in the false vacuum background.

FIG. 16. Penrose diagram corresponding to tunneling along the classically forbidden trajectory of Fig.(3), in the case $1 < \gamma < 2$. In this case the configuration $R \equiv 0$ does not represent a classical solution, and one cannot speak of "bubble nucleation". One possible quantum mechanical interpretation of this process is simply as the birth "from nothing" of a universe composed of two de Sitter domains.

FIG. 17. Integration path $\Gamma_{\epsilon_1, \epsilon_2}$